%% file: ArXivv3.tex
\documentclass[10pt, 
aps,
prx,
twocolumn,
noeprint,
longbibliography,
nofootinbib
]{revtex4-2}
\usepackage{graphicx}
\input{preamble.tex}

\usepackage{color}
\usepackage{cancel}
\begin{document}

\title{Chaotic many-body quantum dynamics, spectral correlations,\\ and energy diffusion}

\author{J.T. Chalker}
\author{Dominik Hahn}

\affiliation{Rudolf Peierls Centre for Theoretical Physics, University of Oxford, Oxford OX1 3PU, United Kingdom}

\date{\today}

\begin{abstract}
We study chaotic many-body quantum dynamics in a minimal model with spatial structure and local interactions. It has a time-independent Hamiltonian, in contrast to quantum circuits and Brownian models, and is simple at the single-site level, in contrast to Sachdev-Ye-Kitaev chains. It is analytically tractable for large local Hilbert space dimension and weak intersite coupling. In this limit we show that energy dynamics is described by a classical master equation and is diffusive. We also show that the spectral form factor can be expressed exactly in terms of the solution to this master equation. For a two-site system we obtain closed-form expressions for both the two-point correlator of energy density and the spectral form factor, in essentially perfect agreement with numerical simulations. For an $L$-site system we show at late times how a linear ramp emerges in the spectral form factor, as universally expected from level repulsion in chaotic quantum systems. Conversely, at earlier times we identify two distinct mechanisms for an increase of the spectral form factor above its ramp value. One of these is associated with energy diffusion and is effective until the Thouless time, which varies as $L^2$. The other involves contributions like those that would appear if the system were composed of many uncoupled subsystems: they generate a large enhancement of the spectral form factor, and are suppressed on a timescale varying as $(\ln L)^2$.  Besides being exact for the limit considered, we believe our approach provides the natural approximation even for small local Hilbert space dimension and strong intersite coupling. We present a numerical study of a spin-half chain, finding an early-time enhancement of the spectral form factor which is qualitatively similar to that in our solvable model.
\end{abstract}
\maketitle

\section{Introduction}\label{sec:introduction}

Generic many-body quantum systems present a theoretical challenge, because no individual example is expected to be analytically tractable, and because numerical studies are limited to small system sizes. An extraordinarily fruitful response to this challenge has been to investigate average properties of ensembles of systems, in place of single instances. In particular, the use of random matrix ensembles to model the Hamiltonian of a generic quantum system has a long and illustrious history~\cite{Mehta_Random_2004,Haake_2010,DAlessio_2016}. In the original and simplest formulation of this approach, the  ensemble is chosen to be invariant under a general, symmetry-respecting rotation of basis. A key diagnostic of the chaotic dynamics expected in generic quantum systems is provided by the statistical properties of the spectrum of the Hamiltonian, and a very wide range of physical systems show universality in their spectral correlations, in the sense that within suitably chosen energy windows these are the same as calculated from random matrix theory. 

Random matrix theory involves two related and well-appreciated simplifications. First, in any fixed basis all matrix elements of the Hamiltonian are non-zero with probability one, whereas for a conventional Hamiltonian consisting only of few-body operators, many matrix elements are zero in a Fock-space basis built from suitable single-particle orbitals. Second, because of invariance of the ensemble under rotations, it cannot fully describe a spatially extended system with local interactions. An important goal is to understand what happens when these unphysical features are removed. 

Efforts have been made in several settings to go beyond these simplifications. Properties of random matrix ensembles that retain the structure following from a restriction to few-body interactions have been studied extensively, as embedded ensembles in nuclear physics~\cite{French_1970,Bohigas_1971,Kota_2001}, and as models for quantum dots in mesoscopic physics~\cite{Blanter_1997,Altshuler_1997,Alhassid_2000}. In addition, in the past decade Sachdev-Ye-Kitaev (SYK) models~\cite{Sachdev_1993,Kitaev_2014} -- fermionic models with random matrix couplings that are quartic or higher order in fermion fields -- have attracted a great deal of attention, partly because of their emergent conformal symmetry at low temperature. Separately, considerable progress has been made recently in understanding the consequences of spatial structure, using either coupled SYK models~\cite{Gu_SYKChain_2017,Davison_2017,Song_2017,Swingle_2022,SYKReview_2022}, or systems with a stochastically time-dependent Hamiltonian, in the forms of random quantum circuits~\cite{NahumQuantumentanglement,vonKeyserlingk_2018,NahumSpreading,Potter_2022,Fisher_Random_2023} or of Brownian unitary time evolution~\cite{Saad_2018,Lamacraft_2018,Xu_2019}. Each of these routes is likely to introduce features that are different from those of a generic, spatially extended, chaotic many-body system: in the first case, because of the exotic low-temperature behaviour at the single-site level, and in the second case, because time-dependence in the Hamiltonian eliminates energy as a conserved density. For generic systems, that are simple at the single-site level and have a time-independent Hamiltonian, while there are important and long-standing general results~\cite{Lieb_1972}, comparable progress has been lacking.

In this paper we combine some of the simplifying features of random matrix theory with spatial structure and local interactions to engineer a tractable model for a chaotic many-body quantum system with a time-independent Hamiltonian. The approach retains one of the motivating ideas from random matrix theory, the notion that it is useful to consider an ensemble of systems and calculate physical properties averaged over this ensemble. At the same time, it necessarily abandons invariance of the ensemble under arbitrary Hilbert space rotations. Instead it uses random matrices as terms in the Hamiltonian acting on single sites or on pairs of sites. As far as we are aware, this is the first example of a time-independent, spatially extended many-body quantum model, with simple properties at the single-site level, for which energy dynamics and spectral correlations can be analysed exactly. 

The model has quantum degrees of freedom supported at each site of a chain, with local Hilbert space dimension $N$, and is parameterised by a dimensionless coupling $\lambda$, which gives the relative strength of two-site to single-site terms. We show for small $\lambda$ and large $N$ that energy dynamics in the model is diffusive and described by a classical master equation, and that statistical properties of spectral correlations of the Hamiltonian can be expressed in terms of the solution to this master equation. 

Our work has links to several recent developments. Most directly, a two-site version of the same model was treated approximately in Ref.~\cite{Altland_2024}, as we discuss further in Sec.~\ref{sec:overview}. More broadly, our approach builds on successes over the past decade in the study of random quantum circuits, which also use random matrices as building blocks to construct models with local couplings \cite{NahumQuantumentanglement,vonKeyserlingk_2018,NahumSpreading,Potter_2022,Fisher_Random_2023}. We find characteristic scales for spectral correlations that have a similar but not identical dependence on system size to those identified previously in random Floquet quantum circuits \cite{Chan_PRX,Chan_PRL,Friedman_2019}. Separately, our calculations offer a microscopic point of comparison for effective field theories of spectral correlations in chaotic quantum many-body systems \cite{Winer_2022,Galitski_2022,altland2025pathintegralapproachquantum}, and we find several common features. 

Models similar to the one we study, in the sense that random matrices are used to represent local interactions of a spatially extended system, have been examined previously from several perspectives, including the form of the density of states \cite{Keating_2015,Collins_2023}, and the validity of the eigenstate thermalisation hypothesis \cite{Sugimoto_2021}. In addition, theoretical simplifications have been discussed in the limit of large local Hilbert space dimension \cite{Laumann_2019}, and for random matrices that square to the identity \cite{Pollock_2025,Pollock2025b}. In the current paper, by restricting ourselves to the limit of weak intersite coupling, we believe we have been able to give a much more extensive treatment of energy transport and spectral correlations than has been attempted in these earlier studies.

The model we study is tractable at large $N$ and small $\lambda$ because in this limit both energy dynamics and spectral correlations can be computed exactly by considering paired Feynman paths in Fock space, in a many-body version of what is known as the diagonal approximation for systems in the semiclassical limit \cite{Berry_85}, and as the diffusion approximation for single-particle models of disordered conductors \cite{Edwards_1958}. Our approach leads directly to a classical master equation for the dynamics of energy density. While we are not able to solve the master equation except for a two-site system, we expect from its form that energy dynamics should be diffusive. More specifically, the symmetry of the master equation means that the nonlinear terms which prevent an analytical solution are irrelevant in the scaling sense. We confirm that dynamics are diffusive using numerical simulations of the master equation, which are enormously simpler than would be the case for the underlying quantum model.

\begin{figure}[t!]
    \centering   \includegraphics[width=0.9\columnwidth]{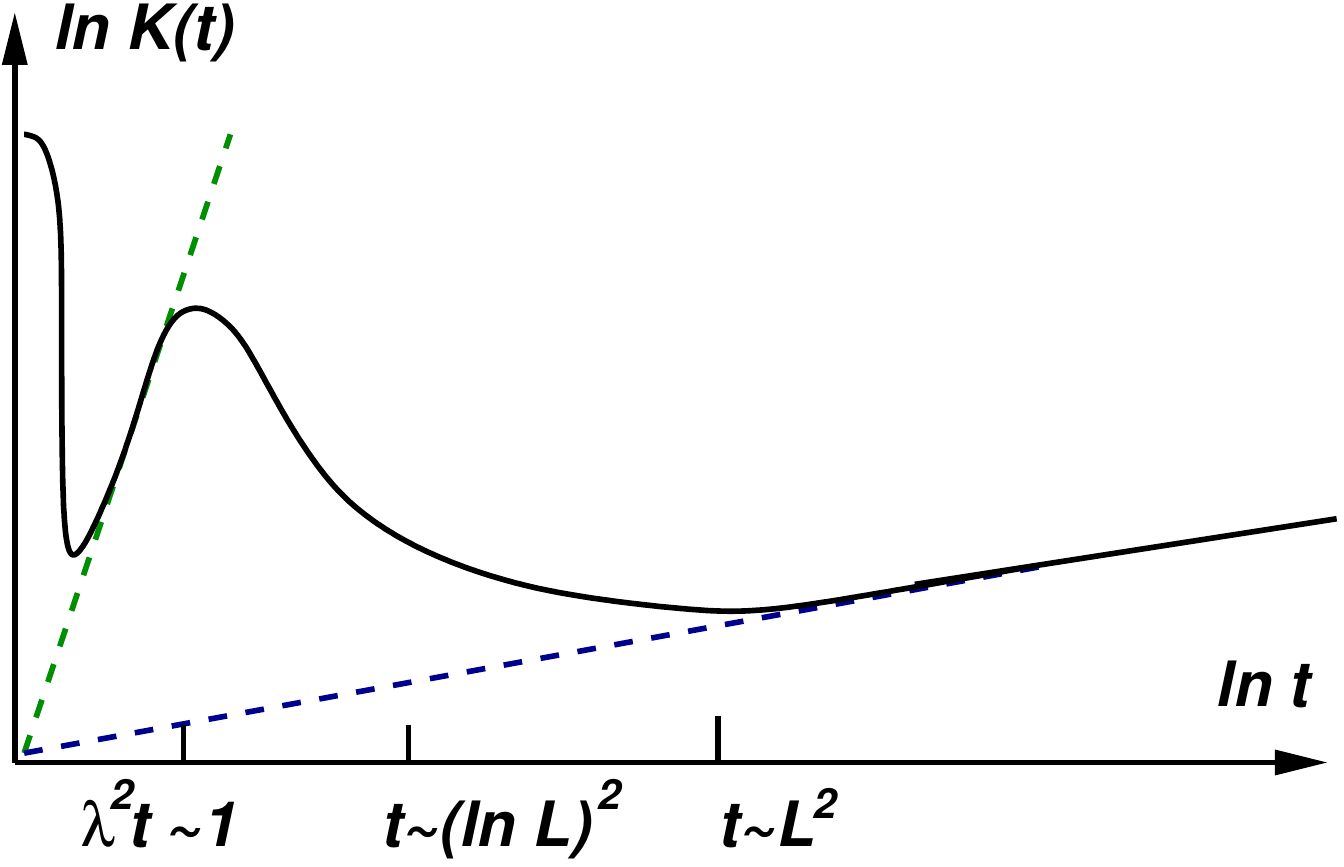}
    \caption{Schematic behaviour of the SFF in a system of $L$ weakly coupled sites, shown as $\ln K(t)$ vs $\ln t$ (black). Dashed lines indicate the behaviours $K(t) \propto t^L$ (green) and $K(t) \propto t$ (blue). See text for discussion of indicated timescales.
    }
    \label{fig:schematic}
\end{figure}

In addition to a characterisation of energy dynamics, the other main objective of our calculations is the spectral form factor (SFF), denoted by $K(t)$, which is the ensemble-averaged Fourier transform of the energy level density, defined in Eq.~\eqref{eq:SFF}. It has a central importance because random matrix features in the SFF are a key diagnostic of chaos in quantum systems~\cite{Mehta_Random_2004,Haake_2010,DAlessio_2016}. Within standard random matrix theory, it has three main features as a function of the transform variable $t$, generally referred to as time. One is a peak centred on $t=0$ with a width in time that is inversely proportional to the energy width of the spectrum. The second is a so-called ramp, with $K(t) \propto t$ at intermediate times, which is a consequence and an indicator of level repulsion. The third is a plateau at times much greater than the inverse level spacing, which is known as the Heisenberg time $t_H$. 

The SFF for the spatially extended model we study shares the peak, ramp and plateau of random matrix theory, but has several additional features, as 
shown schematically in Fig.~\ref{fig:schematic}. At early times, coupling between sites is ineffective and the SFF for a system of $L$ sites is a product of contributions from each site, therefore increasing as $t^L$ after the onset of the single-site ramp. We find that crossover from this behaviour to a ramp linear in time for the many-site system can be understood in a precise way in terms of paired Feynman paths in Fock space. This crossover begins at the scale $\lambda^2 t \sim 1$ but extends over a broad time window if $L$ is large. It depends on two features of the paths. One of these is energy exchange between neighbouring pairs of sites, induced by the coupling; the other is the return probability of paths in Fock space. A ramp appears at times late enough that energy has been exchanged between all neighbouring pairs of sites, and the return probability is independent of time. Both criteria set timescales that grow with system size, as $(\ln L)^2$ and the first case, and as $L^2$ in the second. 

As indicated, these spectral correlations are closely linked to energy dynamics. In more detail, the link is as follows. First, we note that a Feynman path in Fock space that involves energy exchange between some pairs of neighbouring sites, but none between other pairs, amounts to a division of the system into subsystems. In such a division, sites of a given subsystem are linked by energy exchange, while no energy is exchanged between sites from different subsystems. This division has direct consequences for the SFF, since we find that a division of the system into $P$ subsystems contributes to the SFF with a factor $t^P$. As time increases, dominant values of $P$ decrease from $P=L$ at early times ($\lambda^2 t \sim 1 $) to $P=1$ at late times [$\lambda^2 t \sim (\ln L)^2 $]. At still later times ($Dt\sim L^2$), energy diffusion with constant $D$ controls the return probability for the many-body energy density, and the approach of the SFF to a linear ramp. 

Related but not identical behaviour has been found in Floquet quantum circuit models. These show a similar enhancement of the SFF at early times and an approach to the universal ramp at long times, but on a timescale that increases less rapidly with system size~\cite{Chan_PRL}. Floquet circuits in their simplest form lack any conserved density, but variants have been designed with a U(1) symmetry and associated conservation law~\cite{Friedman_2019}. These show show a diffusive timescale in the SFF. Against this background, the work we describe here is intended to identify similarities and differences between spectral correlations in spatially extended, time-independent Hamiltonian models, compared to those in Floquet systems.

While our analysis is exact for the large $N$, small $\lambda$ limit we consider, more generally it is a natural approximation for chaotic many-body lattice models defined by an ensemble for the Hamiltonian. An essential but highly restrictive requirement for any such approximation is that it should respect unitarity for quantum mechanical time evolution by conserving probability density for the many-body wavefunction. Our approach meets this requirement, which may leave little scope for alternative approximations.

The remainder of the paper is organised as follows. In Sec.~\ref{sec:overview} we describe our main results more fully, outline our calculations and present a comparison with numerical calculations for a two-site system. In Sec.~\ref{sec:calculations} we set out details of calculations starting from the quantum many-body Hamiltonian. One result from these calculations is a classical master equation for energy dynamics, and in Sec.~\ref{sec:mastersimulations} we discuss behaviour of solutions to this master equation. In Sec.~\ref{sec:N=2} we present numerical results for a model with $N=2$, showing qualitatively similar phenomena to the ones we derive analytically at large $N$. Finally, in Sec.~\ref{sec:summary} we discuss implications and possible extensions of our work.

\section{Overview}\label{sec:overview}

We define the model we study and the quantities we calculate in Sec.~\ref{sec:model}, and outline our calculations in Sec.~\ref{sec:approach}. A central quantity in our treatment is the transition probability between many-body basis states, and in Sec.~\ref{sec:decomposing} we discuss how it can be separated into multiple contributions according to the extent of energy exchange. We relate this transition probability to energy dynamics and the SFF in Sec.~\ref{sec:SFFIIA}. We present results for a two-site system in Sec.~\ref{sec:twositesIIA} and discuss behaviour in a many-site system in Sec.~\ref{sec:manysites}. We outline some extensions of our main results to systems at finite temperature or with translation invariance in Sec.~\ref{sec:extensions}. We close this overview with a discussion in Sec.~\ref{sec:diagonal} of the use of paired Feynman paths in calculations for chaotic quantum systems of different types, showing how our approach and the form of our results for spectral correlations are related to earlier calculations in the semiclassical limit, in diffusive conductors, and in quantum circuits.

\subsection{Model and definitions}\label{sec:model}

The model we study consists of a chain of $L$ sites with a local Hilbert space of dimension $N$ at each site. The Hamiltonian is
\begin{equation}\label{eq:Hamiltonian}
    {\cal H} \equiv {\cal H}_0 + \lambda {\cal H}_1 = \sum_n H_n + \lambda \sum_n T_{n,n+1}\,,
\end{equation}
where $H_n$ is an $N\times N$ matrix acting at site $n$ (so that the notation is shorthand for $\ldots {\mathds{1}} \otimes {\mathds{1}} \otimes H_n \otimes {\mathds{1}} \otimes \ldots$ with $\mathds{1}$ denoting the $N\times N$ unit matrix) and $T_{n,n+1}$ is an $N^2\times N^2$ matrix coupling sites $n$ and $n+1$. These matrices are chosen independently for each $n$ from the Gaussian unitary ensemble, with zero mean and variances
\begin{equation}\label{eq:Hcorrelator}
    \big[[H_m]_{ij}[H_n]_{kl}\big]_{\rm av} = N^{-1}\delta_{mn}\delta_{il}\delta_{jk}
\end{equation}
and
\begin{equation}\label{eq:Tcorrelator}
    \big[[T_{m,m+1}]_{ij}[T_{n,n+1}]_{kl}\big]_{\rm av} = N^{-2}\delta_{mn}\delta_{il}\delta_{jk}\,,
\end{equation}
where $[\ldots]_{\rm av}$ denotes the ensemble average. Note that Eq.~\eqref{eq:Hcorrelator} sets the energy and time scales (we take $\hbar=1$). Our calculations are exact in the limit $N\to \infty$ and $\lambda\to 0$ with $\lambda \gg N^{-1}$. 

Our objectives are to evaluate the SFF, defined by
\begin{equation}\label{eq:SFF}
    K(t) \equiv \big[\big|{\rm Tr}\, e^{-i{\cal H}t}\big|^2\big]_{\rm av}
\end{equation}
and the two-point correlation function of energy density
\begin{equation}\label{eq:energycorrelator}
    C_{mn}(t) \equiv \big[\langle H_m(t) H_{n} \rangle \big]_{\rm av}
\end{equation}
where $\langle \ldots \rangle \equiv N^{-L} {\rm Tr} \ldots$ is the thermal average in the high temperature limit (see Secs.~\ref{sec:extensions} and \ref{sec:finiteT} for a discussion of finite temperature) and $H_m(t) = e^{i{\cal H}t} H_m e^{-i{\cal H}t}$. For both quantities the time scale of interest is $t \sim \lambda^{-2}$. Our calculations give access to times on this scale but (as is normal for approaches based on resummed perturbation theory \cite{Efetov_1997}) not to times of order $N^L$, the Heisenberg time for the system.

The approach we take closely parallels methods that are standard for single-particle quantum problems with random Hamiltonians, such as models of disordered conductors \cite{Altshuler_85,Akkermans_2007}. Our exact treatment of the many-body system with weak intersite coupling is similar to the diffusion approximation applied to conductors with weak disorder. For disordered conductors, one calculates ensemble-averaged one-particle and two-particle Green functions, generally working in the energy domain. In our context we choose instead to work in the time domain, and the analogue of the two-particle Green function is a disorder-averaged matrix element of $e^{i{\cal H}t} \otimes e^{-i{\cal H}t}$. 

At weak intersite coupling it is natural to work initially in a product basis of eigenstates of the single-site contributions to $\cal H$. We write these basis states as $|\{\nu_n\}\rangle$. They satisfy
\begin{equation}\label{eq:basis}
    H_m|\{\nu_n\}\rangle = \nu_m|\{\nu_n\}\rangle\,.
\end{equation}
Here, to avoid clutter, we omit a label on $\nu_m$ that would distinguish different eigenvalues of $H_m$.

The scaling with $N$ given in Eq.~\eqref{eq:Hcorrelator} ensures that $H_m$ has a finite bandwidth in the large $N$ limit, with the single-site density of states \cite{Mehta_Random_2004,Haake_2010}
\begin{multline}\label{eq:rho}
    \rho(\varepsilon) \equiv \lim_{N \to \infty} N^{-1} [{\rm Tr} \, \delta(\varepsilon - H_n)]_{\rm av}\\
    = \left\{\begin{array}{ccc} (2\pi)^{-1}\sqrt{4-\varepsilon^2}&\quad
         &  \varepsilon^2 \leq 4 \phantom{.}\\ 0&
         & \varepsilon^2 > 4\,.
    \end{array} \right.
\end{multline}
In the small $\lambda$ limit, the total density of states is simply the convolution of the single-site densities, and given at large $N$ by
\begin{multline}\label{eq:rhotot}
    \rho_{\rm tot}(\varepsilon) \equiv \lim_{N\to\infty}N^{-L} \left[{\rm Tr}\, \delta(\varepsilon - {\cal H})\right]_{\rm av} %\\= \int_{-2}^2 {\rm d} \varepsilon_1 \ldots \int_{-2}^2 {\rm d} \varepsilon_L \,\\
    \\= \int {\rm d} \varepsilon_1 \ldots \int {\rm d} \varepsilon_L \,\\
    \rho(\varepsilon_1) \ldots \rho(\varepsilon_L)\, \delta(\varepsilon - \sum_n \varepsilon_n)\,.
\end{multline}
It is zero outside the band edges at $\varepsilon= \pm 2L$ and for large $L$ most of its weight is well approximated by a Gaussian with standard deviation $\sqrt{L}$.

\subsection{Approach to calculations}\label{sec:approach}

We require two combinations of matrix elements in this basis. To calculate $K(t)$ we need 
\begin{equation}\label{eq:Kmatrixelement}
    \langle \{\nu_n\}|e^{-i{\cal H}t}| \{\nu_n\}\rangle \langle \{\nu^{\prime}_n\}|e^{i{\cal H}t}| \{\nu_n^{\prime}\}\rangle\,.
\end{equation}
On the other hand, to calculate $C_{mn}(t)$ we need the transition probability
\begin{equation}\label{eq:Pmatrixelement}
    |\langle \{\nu_n\}|e^{-i{\cal H}t}| \{\nu_n^{\prime}\}\rangle|^2\,.
\end{equation}
Despite the difference between these two expressions, we find in the small $\lambda$, large $N$ limit, that $K(t)$ can be written in terms of contributions to the transition probability. Moreover, we show that the transition probability satisfies a classical master equation.

It is convenient to average the transition probability over realisations of the single-site terms $H_m$ by defining 
\begin{multline}\label{eq:P}
    P(\{\varepsilon_n\},\{\varepsilon^\prime_n\},t) = N^{-L}\Big[ \sum_{\{ \nu_n\}}\sum_{\{\nu_n^\prime\}} 
    \prod_m\delta(\varepsilon_m - \nu_m)\\ \times \delta(\varepsilon^\prime_m - \nu^\prime_m)
    |\langle \{\nu_n\}|e^{-i{\cal H}t}| \{\nu^\prime_n\}\rangle |^2\,\Big]_{\rm av}.
\end{multline}
Here $\nu_m$ and $\nu^\prime_m$ are eigenvalues of a particular realisation of $H_m$, while $\varepsilon_m$ and $\varepsilon^\prime_m$ are fixed energies, independent of the realisation. The factor of $N^{-L}$ ensures that $P(\{\varepsilon_n\},\{\varepsilon^\prime_n\},t)$ is independent of $N$ in the large $N$ limit. Note from this definition that $P(\{\varepsilon_n\},\{\varepsilon^\prime_n\},t)$ is real and non-negative, and has the time-independent normalisation
\begin{multline}
    %\int \ldots \int {\rm d}\{ \varepsilon_n\} 
     \int {\rm d\varepsilon_1}\ldots  \int {\rm d} \varepsilon_L
    \,P(\{\varepsilon_n\},\{\varepsilon^\prime_n\},t)
    = \prod_n \rho(\varepsilon^\prime_n)\,.
\end{multline}
Its initial form is
\begin{equation}\label{eq:Pt=0}
   P(\{\varepsilon_n\},\{\varepsilon^\prime_n\},0)= \prod_n  \delta(\varepsilon_n - \varepsilon_n^\prime)\rho(\varepsilon^\prime_n)\,.
\end{equation}

We compute $P(\{\varepsilon_n\},\{\varepsilon^\prime_n\},t)$ by expanding the time evolution operator in powers of $\lambda$, averaging on two-site couplings, and resumming the contributions that survive for $N\gg 1$ and $\lambda \ll 1$ with fixed $\lambda^2 t$. In this way we find (see Sec.~\ref{sec:calculations}) that it evolves according to a classical master equation 
\begin{multline}\label{eq:master}
    \partial_t P(\{\varepsilon_n \},\{ \varepsilon^\prime_n \};t) =\\ 
    - P(\{\varepsilon_n \},\{ \varepsilon^\prime_n \};t) \int {\rm d} \{\nu_n\} W(\{\nu_n\},\{ \varepsilon_n\})\\
    + \int {\rm d} \{\nu_n\} W(\{\varepsilon_n\},\{\nu_n\})P(\{\nu_n\},\{\varepsilon^\prime_n\}) 
\end{multline}
with loss and gain terms arising from the intersite coupling. The transition rate from the state $\{\varepsilon_n\}$ to the state $\{\nu_n\}$ is given by the Fermi golden rule expression
\begin{multline}\label{eq:transitionrate}
    W(\{\nu_n\},\{ \varepsilon_n\}) = 2\pi \lambda^2 \sum_m \prod_{k\not=m,m+1} \delta(\nu_k - \varepsilon_k)\times\\ \times \rho(\nu_m) \rho(\nu_{m+1}) \delta(\nu_m+\nu_{m+1} - \varepsilon_{m} -\varepsilon_{m+1})\,.
\end{multline}
The factor of $\lambda^2$ appearing in Eq~\eqref{eq:transitionrate} shows that $P(\{\varepsilon_n\},\{\varepsilon^\prime_n\},t)$ depends on time only through the scaled variable $\lambda^2t$. 

Information on energy dynamics follows directly from knowledge of $P(\{\varepsilon_n\},\{\varepsilon^\prime_n\},t)$, since 
\begin{equation}\label{eq:energycorrelator2}
    C_{jk}(t) = \int {\rm d}\{\varepsilon_n\} \int {\rm d}\{\varepsilon^\prime_n\} \, P(\{\varepsilon_n\},\{\varepsilon^\prime_n\},t) \, \varepsilon_j \, \varepsilon_k^\prime\,.
\end{equation}
By contrast, the connection between the SFF and  $P(\{\varepsilon_n\},\{\varepsilon^\prime_n\},t)$ is more involved. It depends on the many-body return probability and on whether energy has been exchanged between each neighbouring pair of sites, as we explain in Sec.~\ref{sec:decomposing}.

\subsection{Decomposing the many-body transition probability}\label{sec:decomposing}

Our next step is to examine in detail the form of contributions to the transition probability, and for this purpose it is useful to make a change of coordinates. Since transitions induced by the coupling ${\cal H}_1$ mix basis states that (in the small $\lambda$ limit) have the same total energy, we change variables from the individual site energies $\varepsilon_m$ and take one of the new variables to be the total energy $\varepsilon_{\rm tot}\equiv \sum_n \varepsilon_n$. For the other variables we use the energies $\omega_n$ transferred from site $n$ to site $n+1$ for $1\leq n \leq L-1$, so that $\varepsilon_1 = \varepsilon_1^\prime - \omega_1$, $\varepsilon_n = \varepsilon_n^\prime + \omega_{n-1} - \omega_n$ for $2\leq n < L-1$,  and $\varepsilon_L = \varepsilon_L^\prime + \omega_{L-1}$. (For simplicity, we consider a system with open boundary conditions, so there is no direct transfer of energy between site $1$ and site $L$.) We also define $\varepsilon^\prime_{\rm tot}\equiv \sum_n \varepsilon^\prime_n$. This change of variables has a Jacobian of unity. Using it, we can introduce the probability distribution $M(\{\omega_n\},\{\varepsilon_n^\prime \},t)$ for energy transfers $\{\omega_n\}$ after evolution for time $t$ from an initial state specified by $\{\varepsilon_n^\prime\}$, by writing
\begin{multline}\label{eq:PtoM}
    P(\{\varepsilon_n\},\{\varepsilon^\prime_n\},t) = M(\{\omega_n\},\{\varepsilon_n^\prime \},t) \\\times \delta(\varepsilon_{\rm tot} - \varepsilon^\prime_{\rm tot})\prod_n \rho(\varepsilon_n^\prime)\,.
\end{multline}
The properties of $P(\{\varepsilon_n\},\{\varepsilon^\prime_n\},t)$ imply that $M(\{\omega_n\},\{\varepsilon_n^\prime \},t)$ is real and non-negative, and has the normalisation
\begin{equation}
    \int {\rm d}\omega_1 \ldots \int {\rm d}\omega_{L-1} M(\{\omega_n\},\{\varepsilon_n^\prime \},t) = 1\,
\end{equation}
as required for a probability distribution.

Initially, there have been no energy transfers and from Eq.~\eqref{eq:Pt=0}
\begin{equation}\label{eq:Mt=0}
    M(\{\omega_n\},\{\varepsilon_n^\prime \},0) = \prod_{n=1}^{L-1}\delta(\omega_n)\,.
\end{equation}
Conversely, in the late time limit, energy is spread uniformly over all accessible states. Then $M(\{\omega_n\},\{\varepsilon_n^\prime \},t)$ is most conveniently written in terms of the coordinates $\{\varepsilon_n\}$ and has the form
\begin{equation}\label{eq:Mtinfty}
   \lim_{t\to \infty} M(\{\omega_n\},\{\varepsilon_n^\prime \},t) = [\rho_{\rm tot}(\varepsilon^\prime_{\rm tot})]^{-1} \prod_n \rho(\varepsilon_n)\,.
\end{equation}

At intermediate times, $M(\{\omega_n\},\{\varepsilon_n^\prime \},t)$ can be decomposed into a sum of contributions, each representing the situation in which there has been no energy transfer between specified pairs of neighbouring sites, but non-zero transfer between the remaining pairs. Every such contribution consists of a product of factors $\delta(\omega_k)$ for the pairs of sites $k, k+1$ without energy transfer, multiplying a smooth function of $\omega_m$ for the pairs of sites $m, m+1$ between which there has been transfer. Specifying a contribution in this way constitutes a division of the sites of the system into subsystems, with energy exchange between all neighbouring sites within each subsystem but none between neighbouring sites from different subsystems. We allocate every site to a subsystem, so if there has been no energy exchange between sites $k-1$ and $k$, and none between sites $k$ and $k+1$, then site $k$ is a subsystem consisting of a single site. We use ${X}$ to label the $2^{L-1}$ possible such divisions, and denote by $P_X$ the number of subsystems in the division $X$. To make these definitions explicit, we introduce an indicator function $I_\ell(X)$ for each neighbouring pair of sites $\ell, \ell+1$, taking the value $I_\ell(X)=0$ if there is no energy transfer between sites $\ell$ and $\ell+1$, and the value $I_\ell(X)=1$ otherwise. This implies $P_{X} = L - \sum_{\ell = 1}^{L-1} I_\ell(X)$. 
Then we write
\begin{eqnarray}\label{eq:Minit}
    M(\{\omega_n\},\{\varepsilon_n^\prime \},t) &=& \sum_X M_X(\{\omega_m\},\{\varepsilon^\prime_n \},t)\times \nonumber\\
    &&\times \prod_{\{k|I_k(X)=0\}} \delta(\omega_k)\,.
\end{eqnarray}
Here the arguments $\{\omega_m\}$ of $M_X(\{\omega_m\},\{\varepsilon_n^\prime \},t)$ are those for which $I_m(X)=1$, and $\prod_{\{k|I_k(X)=0\}}\delta(\omega_k)$ includes only the $\omega_k$ for which $I_k(X)=0$. The quantity $M_X(\{\omega_m\},\{\varepsilon_n^\prime \},t)$ gives the probability density for the joint event that there has been no energy transfer between the set of neighbours $k,k+1$ for which $I_k(X)=0$, and that the energy transfers between the complementary set of neighbours (for which $I_m(X)=1$) take the values $\{\omega_m \}$.

\subsection{SFF and energy dynamics from the many-body transition probability}\label{sec:SFFIIA}

With this background, we can state one of the central result of this paper, which is an expression for the SFF in terms of quantities determined by the classical master equation [Eq.~\eqref{eq:master}] describing energy dynamics. We give separate results for early and late times, which however have an overlapping range of applicability. At early times ($\lambda^2t \ll 1$) there has been no energy exchange between sites, and so 
\begin{equation}\label{eq:SFFearly}
    K(t) = [K_1(t)]^L
\end{equation}
where
\begin{equation}
    K_1(t) = [|{\rm Tr}\, e^{-iH_mt}|^2]_{\rm av}
\end{equation}
is the SFF for a single site. 
At late times (after the onset of the ramp in the single-site SFF) we find %for $t\gg1$
\begin{equation}\label{eq:SFFdecomp}
    K(t) = \sum_X \,\left(\frac{t}{2\pi}\right)^{P_X}\, R_X(t)
\end{equation}
with 
\begin{equation}\label{eq:returnprob}
R_X(t) = \int {\rm d}\{ \varepsilon^\prime_n\} M_X(\{ \omega_m =0\},\{ \varepsilon^\prime_n\},t) \,.
\end{equation}

The expression for the SFF given in Eq.~\eqref{eq:SFFdecomp} has a straightforward physical interpretation. After the onset time for the ramp in the single-site SFF, the many-site SFF consists of a sum of contributions. Each term in this sum is associated with a decomposition $X$ of the full system into $P_X$ subsystems. The term arising from a given decomposition is proportional to $t^{P_X}$ multiplied by a time-dependent weight $R_X(t)$. With increasing time, we expect this weight to shift from being initially concentrated on decompositions $X$ that consist of many small subsystems and therefore have large values of $P_X$, to decompositions at later times that are made up of fewer, larger subsystems and therefore have smaller values of $P_X$. At very late times, all weight is on a single decomposition, which has $P_X=1$ and is equivalent to the full system.

Note that $M_X(\{\omega_n = 0\},\{\varepsilon^\prime_n\})$, appearing in Eq.~\eqref{eq:returnprob}, is the probability for the many-body energy density to return to its initial configuration, given that no energy has been transferred between the pairs of sites that define the decomposition $X$ of the system into subsystems. This feature of our results closely parallels the appearance of the return probability in expressions for the SFF derived for low dimensional systems in the semiclassical limit, and for single particle models of diffusive conductors \cite{Argaman_93}. At the same time, our results for the many-body system include a feature not present in these earlier expressions for the SFF in single-particle models: a large enhancement due to the factor $t^{P_X}$ appearing in Eq.~\eqref{eq:SFFdecomp} in the intermediate time window for which the dominant system divisions $X$ consist of multiple subsystems. 

In Eq.~\eqref{eq:SFFdecomp} we have made simplifications that apply provided $\lambda^2 t\gg 1$ at the Heisenberg time $t \sim N$ for a single site. We give a generalised version of Eq.~\eqref{eq:SFFdecomp} in Eq.~\eqref{eq:fully}. The generalised version avoids these simplifications and holds even when this condition is not satisfied. It is useful when we compare our analytical results with numerical simulations, because it extends the range of $N$ and $\lambda$ for which the predictions apply.  

While Eq.~\eqref{eq:master} is written as an evolution equation for the transition probability $P(\{\varepsilon_k \},\{ \varepsilon^\prime_k \};t)$, it contains full information on the decomposition into separate contributions that we use in our expression for the SFF, Eq.~\eqref{eq:SFFdecomp}. Specifically,  $M_X(\{\omega_n\},\{\varepsilon_n^\prime\},t)$ can be obtained by conditioning this evolution on whether or not there is energy exchange between each neighbouring pair of sites. 

To make use of the correspondence between contributions to the SFF and contributions to the transition probability, we require a solution to the master equation. We derive this analytically for a system of two sites in Sec.~\ref{sec:twosites}. For a system of many sites we do not anticipate that analytic solution is possible. However, we expect on general grounds that behaviour at long times and distances is represented by a diffusion equation with current noise, and in Sec.~\ref{sec:mastersimulations} we use numerical simulations of Eq.~\eqref{eq:master} to demonstrate that this is the case. In this way we can develop a quite detailed idea of the form of the SFF in a system of many sites, as we describe in Sec.~\ref{sec:manysites}.

\subsection{Results for two-site system}\label{sec:twositesIIA}

\begin{figure*}[t!]
    \centering   \includegraphics[width=0.9\textwidth]{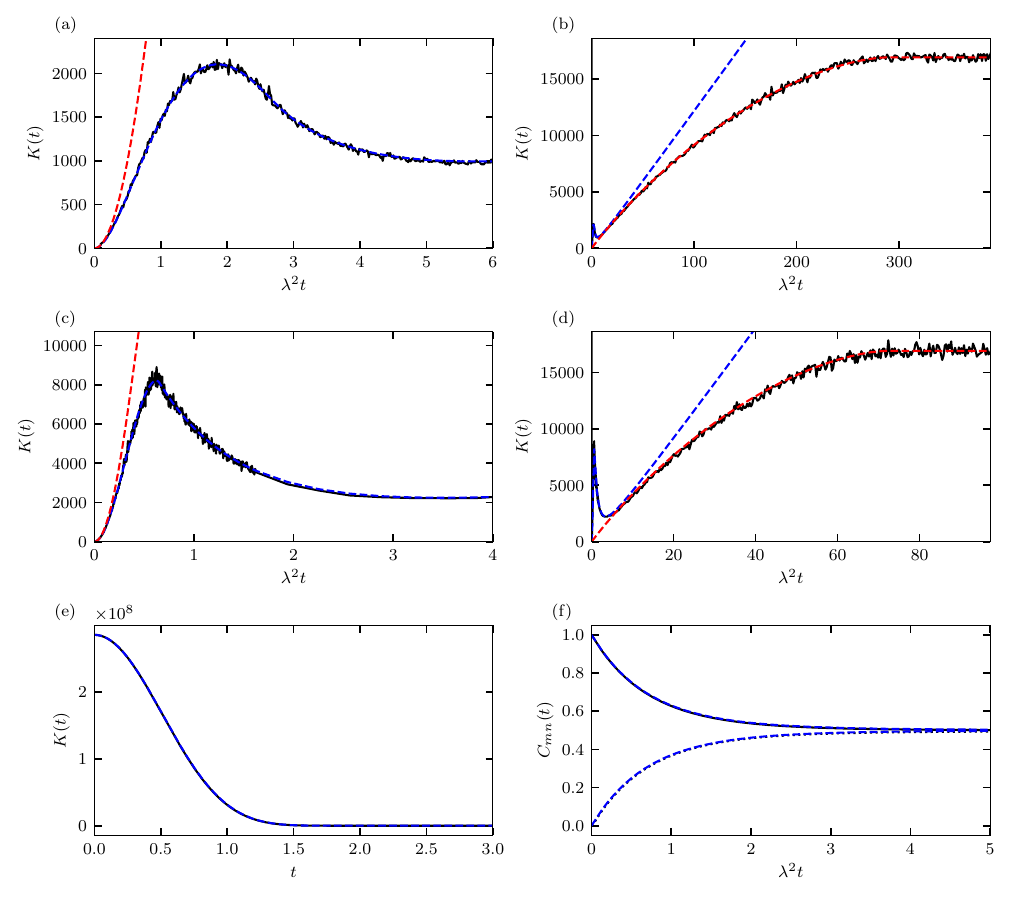}
    \caption{Comparison between numerical and analytical results for the spectral form factor $K(t)$ and the two-point correlation function of energy density $C_{mn}(t)$ with $N=130$ and $L=2$. Intersite coupling strengths are $\lambda=0.1$ for panels (a) and (b), and $\lambda = 0.05$ for panels (c) and (d); panels (e) and (f) are independent of $\lambda$ provided it is small. 
    Panels (a) and (c) show the intermediate-time regime for $K(t)$: numerical data (black), analytical prediction~(blue), and the short-time form $K_1(t)^2$~(red). The analytical results are obtained by adding the contributions $K^{(1)}(t)$ and $K^{(2)}(t)$, Eqns.~\eqref{eq:K^(1)} and \eqref{eq:K^(2)} respectively.
    Panels (b) and (d) show the long-time regime for $K(t)$: numerical data (black), a linear ramp with gradient fixed by the late-time form of Eq.~\eqref{eq:K^(1)} (blue), and the analytical prediction taking into account the shape of the density of states [Eq.~\eqref{eq:K^(1)generalised}] (red).
    (e) $K(t)$ at short times: numerical results~(black) and comparison with analytics~(blue). 
    (f) Comparison between numerical results~(black) and analytical predictions [Eq.~\eqref{eq:predC11}]~(blue) for $C_{11}(t)$~(solid) and $C_{12}(t)$~(dashed).
    All numerical data are averaged over 5000 random realizations.
}
    \label{fig:NumericsSFF}
\end{figure*}

As a first test of our conclusions, we compare in Fig.~\ref{fig:NumericsSFF} our analytical results for the SFF and for energy dynamics in a two-site system (see Sec.~\ref{sec:twosites} for details) with numerical results obtained using exact diagonalisation of the full Hamiltonian. 
Results for the SFF show the features discussed above. At early times [panel (e) of Fig.~\ref{fig:NumericsSFF}] $K(t)= [K_1(t)]^2$, in accordance with Eq.~\eqref{eq:SFFearly}, and at late times [panels (b) and (d) of Fig.~\ref{fig:NumericsSFF}] the SFF shows the ramp and plateau of a generic chaotic quantum system. Between these two limiting regimes there is an intermediate-time peak in the SFF [panels (a) and (c) of Fig.~\ref{fig:NumericsSFF}], dependent on the strength $\lambda$ of intersite coupling. The two-point correlation function of energy density $C_{mn}(t)$ is shown in panel (f) of Fig.~\ref{fig:NumericsSFF}. Energy exchange between sites leads to equilibration at long times. Unlike the SFF, this correlator is a function only of the combined variable $\lambda^2 t$.

Considering these results in more detail, our analytical form for the SFF reduces to the square of the single-site SFF for $\lambda^2t\ll 1$, and gives a ramp with $K(t)= (4/\pi)\, t$ for $\lambda^2 t \gg 1$. At still longer times, $t \sim t_{\rm H} \sim N^{2}$, one expects a crossover from a ramp to a plateau. This crossover is not captured by the calculations we have summarised so far. However, its shape is fixed in a simple way by the density of states for the system, given in Eq.~\eqref{eq:rhotot}. We derive the functional form of the crossover in Appendix~\ref{sec:crossover}: taking it into account, the match between analytical and numerical results is essentially perfect. There is likewise excellent agreement in results for the two-point correlation function $C_{mn}(t)$ of energy densities at each site.

A qualitatively similar approach to the two-site problem has been described in Ref.~\cite{Altland_2024} but with an approximation that introduces two phenomenological parameters, and without a link between the SFF and a master equation, or a discussion of the many-site system. In our terms, the approximation of Ref.~\cite{Altland_2024} amounts to taking the transition rate $W(\{\nu_k\},\{ \varepsilon_k\})$ [Eq.~\eqref{eq:transitionrate}] to be independent of the energies $\{\nu_k\}$ and $\{ \varepsilon_k\}$. While this is an initially appealing simplification, we believe that energy-dependent transition rates are a unavoidable feature of the dynamics, necessary to constrain the energy $\varepsilon_k$ at each site $k$ to lie within the support of the single-site density of states $\rho(\varepsilon_k)$. With the best choice for the values of the phenomenological parameters, the approximate results of Ref.~\cite{Altland_2024} fit simulations well but not exactly.

\subsection{Results for many-site systems}\label{sec:manysites}

Turning to a system of many sites, although our results are less detailed than for two sites, we believe they nevertheless give a rather full picture of the regimes of behaviour for the SFF, and of energy dynamics at large times and distances. 

First, our derivation of a master equation [Eq.~\eqref{eq:master}] for transition probabilities, together with numerical results obtained from this equation and presented in Sec.~\ref{sec:mastersimulations}, shows that the dynamics of energy density is diffusive in our model. While this conclusion is expected, we are not aware of a comparably controlled and detailed calculation of energy transport in a chaotic many-body quantum system. 

Second, a striking sequence of regimes in the behaviour of the SFF as a function of time is exposed by the links we have established between the master equation and the SFF, together with plausible assumptions about the behaviour of the transition probability. We now discuss this, considering in turn the early time ($\lambda^2t \ll 1$), late time ($\lambda^2 t \gg 1$) and intermediate time ($\lambda^2t \sim 1$) regimes. 

At early times ($\lambda^2 t \ll 1$), Eq.~\eqref{eq:SFFearly} applies. Within this initial regime, $K(t)$ first decreases from its maximum value of $N^{2L}$ at $t=0$ to a minimum at the onset time for the ramp in the single-site SFF. We denote this onset time by $t_{\rm dip}$. It scales with $N$ as $t_{\rm dip}\sim N^{1/2}$.  The ramp in $K_1(t)$ extends from the onset time to the Heisenberg time $t_{\rm H} \sim N$ and so for $t_{\rm dip} \ll t \ll t_{\rm H}$, $K_1(t) = 2t/\pi$. 
%with $K(t_{\rm dip}) \sim N^{L/2}$.  
In this time window Eq.~\eqref{eq:SFFearly} gives $K(t)=(2t/\pi)^L$. 

At times after $t_{\rm dip}$ Eq.~\eqref{eq:SFFdecomp} applies. There is overlap between the early time regime and the later time regime if $\lambda\ll N^{-1/4}$, and we assume this to be the case. Our results for the two regimes coincide in the time window where they both apply. To show this, we examine the predictions from Eq.~\eqref{eq:SFFdecomp} at the start of the window of overlap. At this time, no energy transfer has occurred and so all weight is in the decomposition $X$ consisting exclusively of single sites. For this decomposition, represented in Eq.~\eqref{eq:Mt=0}, $P_X=L$ and $M_X(\{\varepsilon_n^\prime\},t)$ has no arguments $\{\omega_m\}$ since there has been no energy transfer. In this case, the value of $R_X(t)$ is the probability for there to be no energy transfer, integrated over the energies of all initial states. The probability is unity at early times and the energy integration range at each site is $(-2,2)$. The value of the integral is therefore $4^L$ In consequence, we find $K(t) = (2t/\pi)^L$, as advertised. 

Next consider very late times ($\lambda^2 t\gg 1$). In this regime there has been energy transfer between all neighbouring pairs and all weight is in the decomposition $X$ for which every site belongs to the same subsystem. For this $X$, $P_X=1$. In addition, at late times energy is uniformly distributed over all accessible states, as represented in Eq.~\eqref{eq:Mtinfty}; setting $\omega_m =0$ in that equation gives
\begin{multline}\label{eq:Rlongt}
    \lim_{t\to \infty} R_X(t) = \int {\rm d}\{\varepsilon^\prime_n\} \,[\rho_{\rm tot}(\varepsilon^\prime_{\rm tot}) ]^{-1}\prod_n \rho(\varepsilon^\prime_n)\\
    =\int {\rm d}\varepsilon_{\rm tot}^\prime\, [\rho_{\rm tot}(\varepsilon^\prime_{\rm tot}) ]^{-1} \times \rho_{\rm tot}(\varepsilon^\prime_{\rm tot})    =4L\,.
\end{multline}
In the second line of Eq.~\eqref{eq:Rlongt} we evaluate the integral by changing coordinates from the $L$ variables $\{ \varepsilon^\prime_n\}$ to the variable $\varepsilon^\prime_{\rm tot}$ and $L-1$ orthogonal coordinates, which we do not specify explicitly; integration on these gives the factor $\rho_{\rm tot}(\varepsilon^\prime_{\rm tot})$, and integration on $\varepsilon^\prime_{\rm tot}$ gives the bandwidth of $\rho_{\rm tot}(\varepsilon^\prime_{\rm tot})$, which is $4L$. In this regime the SFF therefore displays a ramp with $K(t) = (2L/\pi)\,t$.

 The approach of $R_X(t)$ to the long-time value given in Eq.~\eqref{eq:Rlongt} is controlled by the time-dependence of the return probability $M_X(\{\omega_m=0\},\{\varepsilon^\prime_n\},t)$ for the many-body energy density. Taking energy dynamics to be diffusive, we show in Sec.~\ref{sec:masteranalytical} that $R_X(t)$ approaches its long-time value from above, on a timescale $L^2/D$, where $D$ is the energy diffusion constant. %This sets the Thouless time for the system.

In the intermediate time window, $\lambda^2t \sim 1$, the dominant decompositions $X$ contributing to Eq.~\eqref{eq:SFFdecomp} consist of multiple subsystems, so that $P_X\gg1$ if $L$ is large. For this reason, $K(t)$ is strongly enhanced in comparison to its form within the long-time ramp. To model the behaviour of $K(t)$ in this regime, we construct an approximation to $R_X(t)$ following similar ideas to those used the derivation of Eq.~\eqref{eq:Rlongt}. As we now describe, this involves several steps. 

First we introduce the probability ${\cal P}_X(t)$ for the decomposition $X$ to appear at time $t$, averaged over initial conditions. This is given by
\begin{equation}\label{eq:P_X}
    {\cal P}_X(t) = \int {\rm d}\{\omega_m\} {\rm d}\{\varepsilon_n^\prime\} M_X(\{\omega_m\},\{\varepsilon_n^\prime\},t) \prod_n \rho(\varepsilon_n^\prime)\,.
\end{equation}
Second, by neglecting the dependence  on $\omega_m$ of $M_X(\{\omega_m\},\{\varepsilon_n^\prime\},t)$, we can approximate the integrand of Eq.~\eqref{eq:returnprob} as 
\begin{equation}\label{eq:Mapprox}
M_X(\{\omega_m=0\},\{\varepsilon_n^\prime\},t) \approx [\rho_X(\{\varepsilon_n^\prime\})]^{-1} {\cal P}_X(t)\,,
\end{equation}
where $\rho_X(\{\varepsilon_n^\prime\})$ is the integration volume of the integral on $\{\omega_m\}$ in Eq.~\eqref{eq:P_X}. Substituting \eqref{eq:Mapprox} into Eq.~\eqref{eq:returnprob} we have
\begin{equation}\label{eq:Rapprox1}
    R_X(t) \approx {\cal P}_X(t) \int {\rm d}\{\varepsilon^\prime_n\}\, [\rho_X(\{\varepsilon_n^\prime\})]^{-1}\,.
\end{equation}
Now we note that $\rho_X(\{\varepsilon_n^\prime\})$ depends only on the total energy $\varepsilon_{\rm sub}$ of each subsystem in the decomposition $X$, and not on other linear combinations of the $\{\varepsilon_n^\prime\}$. We therefore change integration variables from $\{\varepsilon^\prime_n\}$ to the values of $\varepsilon_{\rm sub}$ for each subsystem and the relative energy coordinates within each subsystem. Integration on these relative coordinates yields $\rho_X(\{\varepsilon_n^\prime\})$, canceling the same factor in the denominator of \eqref{eq:Rapprox1}. Moreover, the integration range for $\varepsilon_{\rm sub}$ in a subsystem of length $L_{\rm sub}$ is $4L_{\rm sub}$. Hence
\begin{equation}\label{eq:Rapprox}
R_X(t) \approx {\cal P}_X \prod_{\rm sub} (4L_{\rm sub})\,,
\end{equation}
where $\prod_{\rm sub}$ runs over the subsystems in the decomposition $X$.

As a final step, we consider ${\cal P}_X(t)$ from a different perspective. We denote by $w(t)$ the probability that there has been no energy exchange between a given neighbouring pair of sites before time $t$, and make the simplifying assumption that this is statistically independent for each pair (numerical evidence presented in Sec.~\ref{sec:masterlate} supports this approximation). The probability for a given decomposition $X$ to occur is then
\begin{equation}\label{eq:Papprox}
    {\cal P}_X(t) = w(t)^{n_X} [1-w(t)]^{L-1-n_X}
\end{equation}
where $n_X$ is the number of neighbouring pairs of sites between which there has been no energy exchange in $X$, and $L-1-n_X$ is the complementary number of neighbouring pairs in an $L$-site system with open boundary conditions. 

Combining Eqns.~\eqref{eq:SFFdecomp}, \eqref{eq:Rapprox} and \eqref{eq:Papprox}, we arrive at an approximation for the SFF
\begin{equation}\label{eq:Kw}
    K(t) \approx a(t) t\{[1-w(t)]+tw(t) \}^{L-1}\,.
\end{equation}
Here, $a(t)$ includes factors of $(2\pi)^{-P_X}$ from Eq.~\eqref{eq:SFFdecomp} and factors of $4L_{\rm sub}$ from Eq.~\eqref{eq:Rapprox}. From the discussion above of the early and late time regimes, the value of $a(t)$ interpolates between $a(t) = (2/\pi)^L$ for $t_{\rm dip} \ll \lambda^2t \ll 1$ and $a(t) = 2L/\pi$ for $\lambda^2 t \gg 1$. While these limiting values of $a(t)$ are widely separated for large $L$, at small $\lambda$ we have $\ln a(t) \ll \ln K(t)$ throughout the intermediate time regime. In this sense, the dominant contribution to $K(t)$ is the factor $t\{[1-w(t)]+tw(t) \}^{L-1}$.

This factor matches the the form found previously for the SFF in Floquet quantum circuits \cite{Chan_PRL}. It can be read as follows. The leading factor of $t$ represents the ramp in the SFF of the subsystem containing (say) the left-most site, and the $L-1$ factors of $[1-w(t)] + t\,w(t)$ are associated with successive neighbouring pairs of sites, from left to right along the length of the system. For each of these, the term $[1-w(t)]$ describes the situation in which energy has been exchanged between the pair of sites, while the term $t\,w(t)$ describes the situation without exchange. In the second case, the term includes a factor of $t$ because the two sites in the pair belong to different subsystems, which each contribute to $K(t)$ with a factor of $t$.

As time increases, $w(t)$ decreases monotonically from unity to zero on the timescale $\lambda^2 t\sim 1$ (see Sec.~\ref{sec:masterlate} for details). Correspondingly, the factor $[1-w(t)] + t\,w(t)$ increases roughly as $t$ until a maximum at $\lambda^2 t \sim 1$, and then decreases towards unity. As a result, $K(t)$ for large $L$ initially increases rapidly with $t$, passes through a peak at $\lambda^2t \sim 1$, and approaches a ramp at long times. The timescale for the crossover from the intermediate-time peak to the ramp is set at large $L$ by $L t w(t) \sim 1$. We find in Sec.~\ref{sec:masterlate} [see Eq.~\eqref{eq:wlate}] that $w(t) \sim A e^{-b\sqrt{\lambda^2t}}$ for late times, where $A$ and $b$ are positive constants, and so the crossover timescale varies with $L$ as $(\ln L)^2$. Note for large $L$ that this is a much shorter time than the Thouless time discussed above, which is not captured by Eq.~\eqref{eq:Kw} since that equation omits the relevant aspect of the time-dependence of $R_X(t)$. 

\subsection{Extensions to the main results}\label{sec:extensions}

Our main results are for a system without translation invariance [since the terms $H_n$ and $T_{n,n+1}$ in Eq.~\eqref{eq:Hamiltonian} are statistically independent for each $n$] and for the high-temperature limit [since this is the average we use in Eq.~\eqref{eq:energycorrelator}]. We have also examined in outline the consequences of relaxing each of the restrictions. 

A translationally invariant version of the model we have studied can be defined by omitting the factors of $\delta_{mn}$ on the right-hand sides of Eqns.~\eqref{eq:Hcorrelator} and \eqref{eq:Tcorrelator}. While a full analysis of this version would present new challenges, two conclusions are straightforward to derive. First, energy dynamics is unchanged at large $N$ and small $\lambda$. Second, in the late-time regime (where, with probability close to one, energy has been transferred between all nearest-neighbour pairs in the system) the spectral from factor of the translationally invariant system is larger than without translational invariance by a factor of $L$ (related results for Floquet circuits have been obtained previously \cite{Chan_2022}). See Sec.~\ref{sec:TranslationInv} for further discussion.

Our studies of energy transport can be extended without difficulty at large $N$ and small $\lambda$ to non-zero inverse temperature $\beta$. We find that the energy diffusion constant is proportional to $\beta^{-2}$ in the low-temperature limit, reflecting the reduced density of states for energy exchange between sites in this regime. See Sec.~\ref{sec:finiteT} for details.

\subsection{Paired Feynman paths and the diagonal approximation}\label{sec:diagonal}

A ramp is present in the SFF for many physical systems, and can be understood using a language first developed in studies of low-dimensional quantum chaos in the semiclassical limit. In that setting, starting from the Feynman path-integral formulation of quantum mechanics, the SFF is expressed as a sum over contributions from pairs of classical periodic orbits. Pairs of orbits that are the same up to a relative translation in time make contributions that interfere constructively, and the ramp is generated by integration over this relative coordinate \cite{Berry_85}. Calculations of energy level correlations for single-electron models of weakly disordered conductors also imply a ramp \cite{Altshuler_86} and can be interpreted in just the same way \cite{Argaman_93}. A similar perspective applies in Floquet quantum circuits, with a ramp in the SFF arising from constructive interference between pairs of Feynman paths \cite{Kos_2018,Chan_PRX,Friedman_2019,Bertini_2018,Chan_PRL,Garratt_2021,Kumar_2025}, where in this case the paths are in Fock space. An approach of this type is known as the diagonal approximation in the semiclassical context, and as the diffusion approximation for disordered conductors.

While these calculations for a variety of systems all yield a ramp in the SFF from interference between paired paths, they each rest on different justifications. For low-dimensional quantum chaos, as noted, the picture applies in the semiclassical limit \cite{Berry_85}, and for weakly disordered conductors the relevant small parameter is the ratio of the Fermi wavelength to the mean free path \cite{Altshuler_86}. In Floquet quantum circuits, the equivalent picture can be justified either by long-range couplings \cite{Kos_2018,Kumar_2025}, or by large local Hilbert space dimension \cite{Chan_PRX,Chan_PRL,Friedman_2019}, or in dual unitary circuits \cite{Bertini_2018}. 

The presence of a ramp in the SFF for diverse systems is an indication of the universality of random-matrix level correlations, which applies beyond a minimum timescale. Like the ramp itself, the timescale for its onset can also be understood from the behaviour of the relevant Feynman paths. In general, it is the earliest timescale after which the probability for paired paths to close on themselves is independent of their duration. The physical origin of this time depends on the setting. In the semiclassical limit, it is given by the period of the shortest periodic orbits \cite{Berry_85}. In weakly disordered conductors it is set by the time taken for an electron to spread diffusively across the entire system; for a single-particle system with density diffusion constant $D$ and linear size $L$, this is $t_{\rm Th}=L^2/D$ and known as the Thouless time \cite{Altshuler_86,Thouless_77}. 

The origin of the onset time for a ramp in the SFF has also been quite widely studied in Floquet quantum circuits, as we now discuss. Two distinct mechanisms have been identified in Floquet quantum circuits that set this timescale according to the circumstances. One of these mechanisms is essentially the same as that operating in single-particle systems, and the associated timescale is the time for paired Feynman paths to explore Fock space uniformly \cite{Kos_2018,Friedman_2019,Garratt_2021,Kumar_2025}. In the presence of a U(1) symmetry and a conserved density, this time is set by diffusion and therefore varies with system size as $L^2$ \cite{Friedman_2019}. In the absence of such symmetry the same mechanism generates a timescale varying as $\ln L$ \cite{Kos_2018}. 

A second mechanism is also important in spatially extended Floquet quantum circuits with local interactions, for which the SFF generically shows large deviations from the ramp at early times. This behaviour can be summarised by first considering a modified circuit in which some gates have been removed, leaving a system that consists of uncoupled pieces. For the modified circuit with $P$ uncoupled pieces, each piece contributes multiplicatively to the SFF, which in place of a linear ramp therefore varies as $t^P$. As a result, the SFF is greatly enhanced for $t>1$ if $P$ is large. The SFF for an unmodified Floquet quantum circuit, in which all sites are coupled to neighbours by gates, has similar behaviour at early times. Crossover to the ramp of random matrix theory occurs at a generalised Thouless time, which is given in a solvable model by $t_{\rm Th} = \varepsilon^{-1} \ln L$, where $\varepsilon$ parameterises the strength of coupling between sites \cite{Chan_PRL,Garratt_2021}. 

The work we present in this paper shows that paired Feynman paths in Fock space give an exact treatment of our model at large $N$ and small $\lambda$. Specifically, it is these contributions that are captured by the master equation that we derive  in Sec.~\ref{sec:approach} for energy dynamics, and via the link between energy dynamics and the SFF set out in Sec.~\ref{sec:SFFIIA} it is the same contributions that control spectral correlations.

\section{Calculations}\label{sec:calculations}

In this section we describe the main steps of the calculations that lead to the results outlined in Sec.~\ref{sec:overview}. In Sec.~\ref{sec:expansion} we derive the master equation for many-body transition probabilities [Eq.~\eqref{eq:master}] by ensemble averaging and resumming the perturbative expansion for the time evolution operator in powers of the intersite coupling; in Sec.~\ref{sec:PtoK} we  derive the relationship [Eq.~\eqref{eq:SFFdecomp}] between contributions to the SFF and the solution to this master equation; in Sec~\ref{sec:TranslationInv} we discuss in outline a translationally invariant model; and in Sec.~\ref{sec:twosites} we apply our general approach to a system of two sites. 

\subsection{Perturbative expansion}\label{sec:expansion}

The starting point for our calculations of the SFF [Eq.~\eqref{eq:SFF}] and the transition probability [Eq.~\eqref{eq:P}] is the expansion of the time evolution operator in the interaction representation 
\begin{eqnarray}\label{eq:interactionrep}
e^{-i{\cal H}t} &=& e^{-i{\cal H}_0t} - i\lambda\int_0^t {\rm d}t_1 \,\,e^{-i{\cal H}_0 (t-t_1)}{\cal H}_1 e^{-i{\cal H}_0 t_1}\nonumber \\&&+ (i\lambda)^2 \int_0^t {\rm d}t_1 \int_0^{t_1} \!\! {\rm d}t_2\,\,\times \nonumber \\
&&\times \, \,e^{-i{\cal H}_0 (t-t_1)}{\cal H}_1 e^{i{\cal H}_0 (t_2-t_1)}{\cal H}_1 e^{-i{\cal H}_0 t_2} \nonumber \\
&&+ \ldots\,.
\end{eqnarray}

\begin{figure}[htb]
    \centering   \includegraphics[width=0.8\columnwidth]{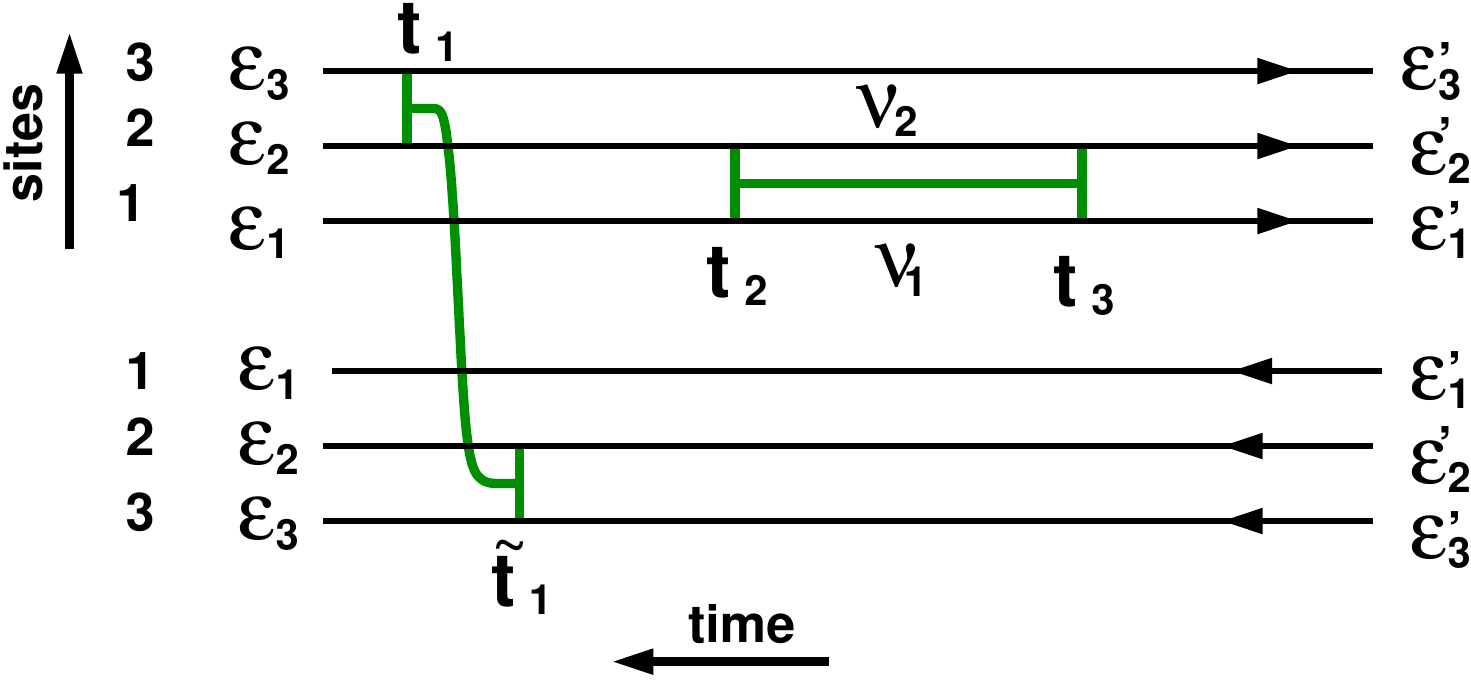}
    \caption{A contribution to $P(\{\varepsilon_k \},\{ \varepsilon^\prime_k \};t)$ at order $\lambda^4$ for a system of three sites. The three horizontal lines directed to the right represent timelines of sites contributing to $e^{-i{\cal H}t}$ and the three lines directed to the left are the equivalent for $e^{i{\cal H}t}$. The green lines represent Wick-paired factors of $T_{n,n+1}$ after the ensemble-average. The Wick pair shown here with time labels $t_1$ and $\tilde{t}_1$ is a vertex contribution, and the one with time labels $t_2$ and $t_3$ is a self-energy contribution. 
}
    \label{fig:diagrams1}
\end{figure}

After the ensemble average on $T_{n,n+1}$, terms in the expansion can be represented diagrammatically as illustrated in Fig.~\ref{fig:diagrams1}. The building blocks of these diagrams are single-site propagators and intersite coupling insertions, which are Wick-paired after averaging. We refer to these Wick pairs as self-energy contributions if both insertions of $T_{n,n+1}$ come from $e^{i{\cal H}t}$ or if both come from $e^{-i{\cal H}t}$, and as vertex contributions if one insertion comes from each exponential. Following Eq.~\eqref{eq:interactionrep}, every insertion carries a time coordinate. Time coordinates from the expansion of a given exponential are ordered, but a pair of time coordinates coming from expansions of different exponentials have no fixed relative order. In addition, working in the basis of eigenstates of ${\cal H}_0$ introduced in Eq.~\eqref{eq:basis}, the propagator for site $m$ carries an energy label $\nu_m$. From Eq.~\eqref{eq:Tcorrelator}, these energy labels are pairwise equal at self-energy and vertex insertions. Contributions are evaluated by associating a factor of $e^{\pm i\nu_m (t^\prime - t^{\prime\prime})}$ with a propagator for site $m$ between end points at times $t^\prime$ and $t^{\prime\prime}$ and a factor of $\lambda^2$ with each paired insertion, summing over internal energies $\nu_m$ and integrating over internal time coordinates. 

Calculations simplify for $\lambda \ll 1$ because in this regime the dominant self energy and vertex contributions are widely spaced in time. More specifically, denote the two time coordinates associated with a given self-energy or vertex contribution by $t_1$ and $t_2$, and change from these variables to the mean and relative times, $(t_1+t_2)/2$ and $t_2 - t_1$ respectively. The contribution is sub-leading unless its weight $\lambda^2$ is offset by an integration window  $\sim \lambda^{-2}$ for its mean time coordinate. This means that the only self-energy and vertex contributions are the lowest order ones illustrated in Fig.~\ref{fig:diagrams1}, which however may each appear multiple times if $\lambda^2 t$ is large. Since contributions are widely spaced in time, integration on relative time coordinates can be extended to cover the range $[0,\infty)$ for self-energy contributions, and $(-\infty,\infty)$ for vertex contributions. 

Terms in the expansion that contribute for $N\gg 1$ have factors of $N^{-2}$ from Eq.~\eqref{eq:Tcorrelator} which are compensated by factors of $N$ appearing from sums over energy labels $\nu_m$. At large $N$, these energy sums can be replaced by integrals weighted with the ensemble-averaged single-site density of states, $\rho(\nu_m)$, given in Eq.~\eqref{eq:rho}. 

Our calculations require $\lambda$ small and $N$ large, with $\lambda \gg N^{-1}$. From a physical viewpoint, in the opposite regime of $\lambda \to 0$ at fixed large $N$ one expects a many-body localised phase and no simple analytical treatment.  At a technical level, this is because integration over the relative time coordinate at a vertex imposes energy conservation to an accuracy $\lambda^2$ on the incoming and outgoing site energies. For a pair of sites $m$ and $m+1$ we require there to be many energies $\nu_m + \nu_{m+1}$ within this window, in order to replace energy sums with integrals, which necessitates $\lambda \gg N^{-1}$. 

\begin{figure*}[t!]
    \centering   \includegraphics[width=1.9\columnwidth]{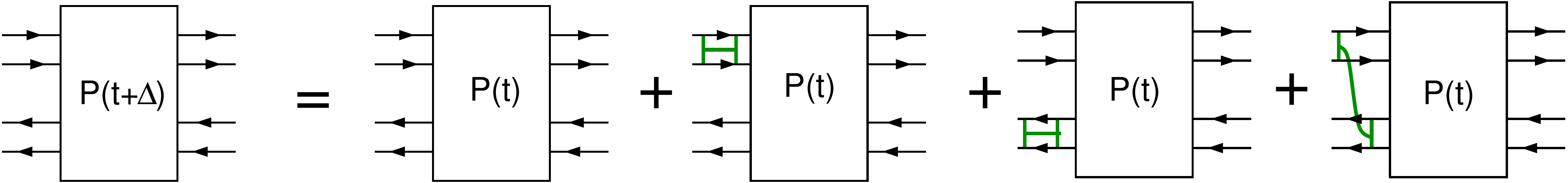}
    \caption{Illustration of the Bethe-Salpeter equation satisfied by the transition probability $P(\{\varepsilon_n \},\{ \varepsilon_n^\prime \};t)$ in the case of a two-site system. The boxes labeled $P(t)$ and $P(t+\Delta)$ represent the transition probability evaluated at times $t$ and $t+\Delta$.
}
    \label{fig:diagrams3}
\end{figure*}

In order to resum the contributions to the transition probability $P(\{\varepsilon_n \},\{ \varepsilon_n^\prime \};t)$ that survive for small $\lambda$ and large $N$, we write a Bethe-Salpeter equation for the evolution of this probability over a time interval $\Delta$, with $\lambda^2 \Delta \ll 1$ but $\Delta\gg 1$. This equation is illustrated in Fig.~\ref{fig:diagrams3}. The second and third terms on the right-hand side of this diagrammatic equation have self-energy insertions in the time interval $[t,t+\Delta]$. Using the simplifications described, they make contributions
\begin{multline}\label{eq:selfenergy}
    - \lambda^2 \Delta \,P(\{\varepsilon_n \},\{ \varepsilon_n^\prime \};t) \sum_m\int_0^\infty {\rm d} t^\prime 
    \\ \times \int {\rm d}\nu_m\int {\rm d}\nu_{m+1}\,
      \rho(\nu_m)\, \rho(\nu_{m+1})\\ \times 
      e^{\pm i (\varepsilon_m + \varepsilon_{m+1} - \nu_m - \nu_{m+1})t^\prime}\,.
\end{multline}
Some of the factors appearing in this expression arise as follows. Integration over the mean time coordinate of the self-energy insertion gives $\Delta$, the integral on $t^\prime$ is over the relative time coordinate of the insertion, and the sum on $m$ appears because in a many-site system this insertion can be located between any pair of neighbouring sites $m$ and $m+1$. All propagators combine to give the factor $e^{\pm i (\varepsilon_m + \varepsilon_{m+1} - \nu_m - \nu_{m+1})t^\prime}$ with a sign determined by whether the self energy is from $e^{i{\cal H}t}$ or  $e^{-i{\cal H}t}$. Finally, $\nu_m$ and $\nu_{m+1}$ are energy labels for propagators within the self-energy insertion; integration over these labels is weighted by single-site densities of states $\rho(\nu_{m})$ and $\rho(\nu_{m+1})$. In a similar way, the fourth term on the right-hand side of the equation shown in Fig.~\ref{fig:diagrams3} has a vertex insertion in the time interval $[t,t+\Delta]$ and makes the contribution
\begin{multline}\label{eq:vertex}
    + \lambda^2 \Delta\sum_m\int_{-\infty}^\infty {\rm d} t^\prime  \int {\rm d}\nu_m\int {\rm d}\nu_{m+1}\\
     \times \rho(\varepsilon_m) \rho(\varepsilon_{m+1})\,P(\{\nu_n \},\{ \varepsilon^\prime_n \};t)\\
     \times e^{ i (\varepsilon_m + \varepsilon_{m+1} - \nu_m - \nu_{m+1})t^\prime}\,.
\end{multline}
These expressions lead directly to the master equation [Eq.~\ref{eq:master}] for the transition probability: the two self-energy contributions [Eq.~\eqref{eq:selfenergy}] give the loss term [the first term on the right-hand side of Eq.~\eqref{eq:master}] and the vertex contribution [Eq.~\eqref{eq:vertex}] gives the gain term [the second term on the right-hand side of Eq.~\eqref{eq:master}]. The initial condition for the transition probability is given in Eq.~\eqref{eq:Pt=0}.
 
\subsection{Relation between $P(\{\varepsilon_n \},\{ \varepsilon^\prime_n \};t)$ and $K(t)$}\label{sec:PtoK}

The next stage in our approach is to establish a term-by-term correspondence between the diagrammatic expansions for the transition probability and for the SFF, which holds in the limit of large $N$ and small $\lambda$. We state the result in Sec.~\ref{sec:correspondence}, then outline a derivation in Secs.~\ref{sec:proof} and \ref{sec:proof2}. An example of the correspondence is shown in Fig.~\ref{fig:KtoP}.

\subsubsection{Correspondence}\label{sec:correspondence}

 Consider a diagram $\cal G$ in the expansion for the transition probability. It defines a decomposition of the system into fully connected subsystems, in analogy with the discussion above Eq.~\eqref{eq:Minit}. In this decomposition, every nearest-neighbour pair of sites within a subsystem is linked by at least one vertex insertion, and there are no vertex insertions between neighbouring sites from different subsystems (the presence or absence of self-energy insertions is not relevant to the decomposition). For example, for the diagram shown in Fig.~\ref{fig:diagrams1} this decomposition consists of two subsystems: one of these is the single site 1 and the other is the pair of sites 2 and 3. As a second example, for the diagram shown in Fig.~\ref{fig:KtoP}(c) there is a single subsystem since the two sites are connected by vertices. 
 
 The contribution of $\cal G$ to $P(\{\varepsilon_n \},\{ \varepsilon^\prime_n \};t)$ can be written in the form 
\begin{multline}\label{eq:PDiagramContribution}
   M_{\cal G}(\{\omega_m \},\{ \varepsilon^\prime_n \};t)\,\prod_k \delta(\omega_k) \times \\ \times
   \delta(\varepsilon_{\rm tot} - \varepsilon^\prime_{\rm tot}) \prod_n \rho(\varepsilon^\prime_n)\,.
\end{multline}
Here the variables $\omega_m$, $\varepsilon_{\rm tot}$ and $\varepsilon^\prime_{\rm tot}$ are defined above Eq.~\eqref{eq:PtoM}. If sites $n$ and $n+1$ lie in the same subsystem, the variable $\omega_n$ appears as an argument of $M_{\cal G}(\{\omega_m \},\{ \varepsilon^\prime_n \};t)$; otherwise it appears in $\prod_k \delta(\omega_k)$.

We find that diagrams for $K(t)$  at large $N$ and small $\lambda$ can be put into a one-to-one correspondence with those for $P(\{\varepsilon_n \},\{ \varepsilon^\prime_n \};t)$. The contribution to $K(t)$ of the diagram corresponding to $\cal G$ depends on the decomposition into subsystems that it defines. Let $T_{\cal G}$ be the number of subsystems in $\cal G$ consisting of two or more sites, and $S_{\cal G}$ the number of single-site subsystems. The contribution to $K(t)$ corresponding to $\cal G$ is
\begin{multline}\label{eq:fully}
    \left(\frac{t}{2\pi}\right)^{T_{\cal G}} \int {\rm d} \{\varepsilon^\prime_n\}\, \times \\ \times M_{\cal G}(\{\omega_m=0 \},\{ \varepsilon^\prime_n \};t)\,\prod_{k}k(\varepsilon_k^\prime,t)\,,
\end{multline}
where $\prod_k$ runs over  single-site subsystems and the function $k(\varepsilon_k^\prime,t)$ is given in Eq.~\eqref{eq:Def-k}. For $t \ll N$ we have $k(\varepsilon_m^\prime,t)= t/(2\pi)$ and Eq.~\eqref{eq:fully} simplifies to
\begin{equation}\label{eq:fully2}
    \left(\frac{t}{2\pi}\right)^{P_{\cal G}} \int {\rm d} \{\varepsilon^\prime_n\}\,  
    M_{\cal G}(\{\omega_n=0\},\{ \varepsilon^\prime_n \};t)\,,
\end{equation}
where $P_{\cal G} = T_{\cal G} + S_{\cal G}$ is the number of subsystems of all sizes in $\cal G$. Summation of this over all diagrams $\cal G$ generates the result for $K(t)$ given in Eq.~\eqref{eq:SFFdecomp}.

\subsubsection{Main ideas: fully connected diagrams}\label{sec:proof}

Consider a diagram $\cal G$ for $P(\{\varepsilon_n \},\{ \varepsilon^\prime_n \};t)$ that is fully connected in the sense defined above, and that contributes at large $N$ and small $\lambda$. Our main point can be illustrated using diagrams without self-energy insertions, and so we start by discussing this case. Let $v$ be the number of vertices. Evaluation of the diagram involves energy integrals (integration over $L+2v$ single-site energies) and time integrals (integration over the mean and relative time coordinates for each vertex). For the example shown in Fig.~\ref{fig:KtoP}(c), $L=2$ and $v=2$.

\begin{figure}[tbh]
    \centering   \includegraphics[width=0.3\columnwidth]{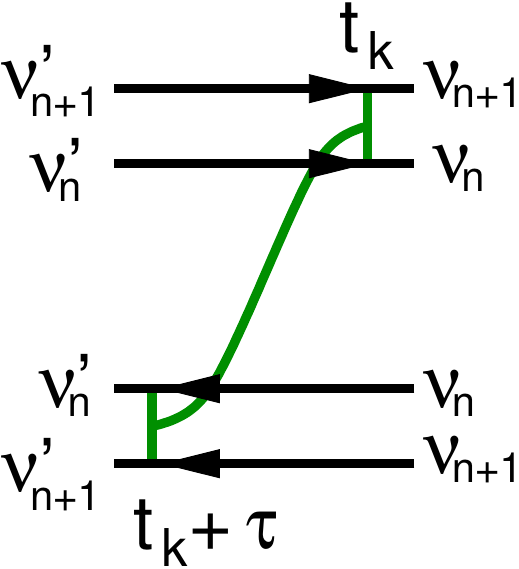}
    \caption{A vertex insertion, labeled as in Eq.~\eqref{eq:energyconservation}. 
}
    \label{fig:vertex}
\end{figure}

\begin{figure}[h!]
    \centering   
    \includegraphics[width=0.6\columnwidth]{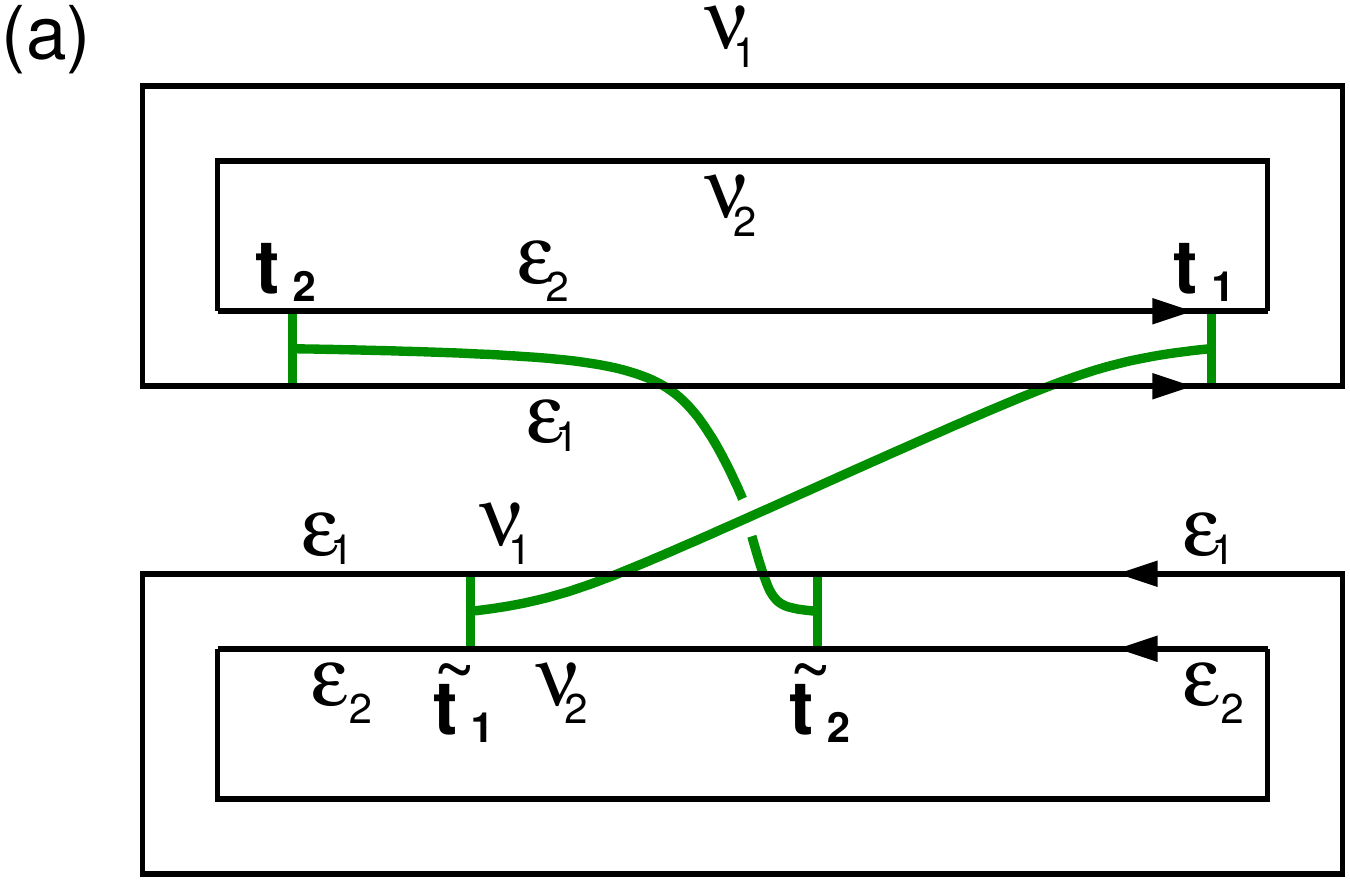}\\ \vspace{0.6cm}
    \includegraphics[width=0.6\columnwidth]{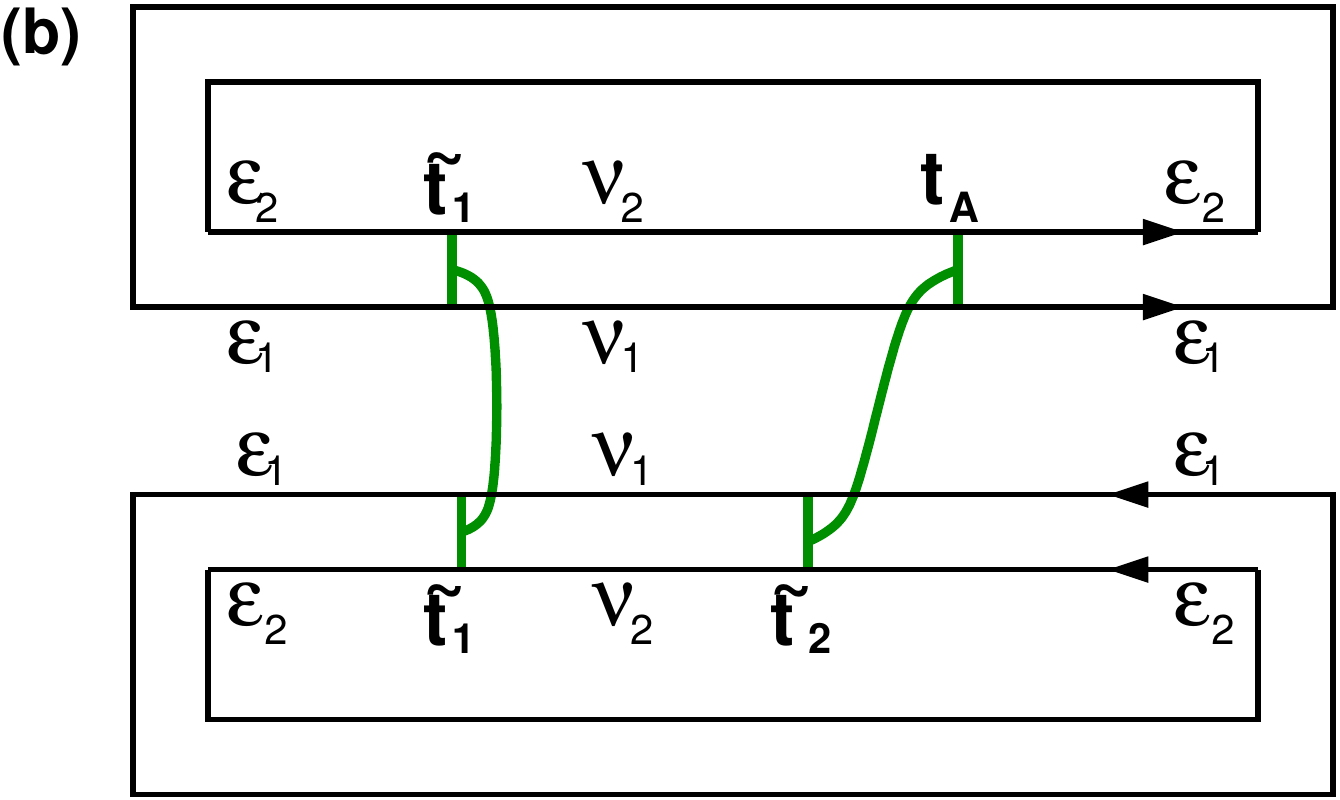}\\ \vspace{0.6cm}
    \includegraphics[width=0.6\columnwidth]{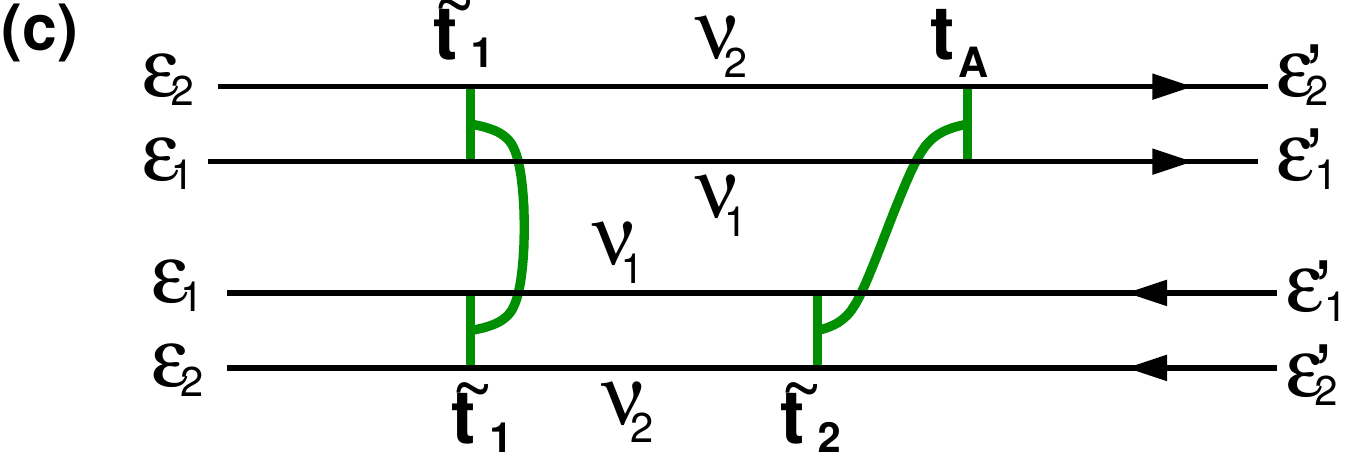}
    \caption{Illustration of the mapping of a diagram for $K(t)$ to one for $P(\{\varepsilon_k \},\{ \varepsilon^\prime_k \};t)$. (a) A diagram for $K(t)$ at order $\lambda^4$ in a system of two sites. (b) A diagram for $K(t)$ obtained from (a) by the transformation $t_1 \to t_1+ \tau$ (modulo $t$) and $t_2 \to t_2+ \tau \equiv t_A$ (modulo $t$), setting $\tau = \tilde{t}_1 - t_1$ (modulo $t$). (c) The contribution to $P(\{\varepsilon_n\},\{\varepsilon^\prime_n\},t)$ that corresponds to (b) under this mapping, after setting $\{\varepsilon_n^\prime\} = \{\varepsilon_n\}$. 
}
    \label{fig:KtoP}
\end{figure}

We start with the relative time integrals. With energy labels for single-site propagators as illustrated in Fig.~\ref{fig:vertex}, integration over the relative time coordinate at one vertex gives
\begin{multline}\label{eq:energyconservation}
    \int_{-\infty}^\infty {\rm d} \tau \, e^{i\tau(\nu_n+\nu_{n+1} - \nu^\prime_n - \nu^\prime_{n+1})}\\ = 2\pi \,\delta(\nu_n+\nu_{n+1} - \nu^\prime_n - \nu^\prime_{n+1})
\end{multline}
and so ensures energy conservation in the scattering process at this vertex. Integrals on the mean times of each vertex are independent of site energies, 
%The energy integrals then reduce to integration over $L$ values for the site energies at $t=0$ and integration over $v$ energy transfers at the vertices. The integrand is independent of the mean time coordinates of the vertices,
%Moreover, the integrals over the mean time coordinates are independent of energies, and 
so that the integrals on mean times and on energies factorise.

Next we examine the energy integrals in more detail. They involve $2v+L$ site energies, which we denote for site $n$ by $\nu_n$, $\nu^\prime_n$, $\nu_n^{\prime\prime} \ldots$. It is convenient to make the following change of variables. We use the values of these energies at time zero as $L$ of our coordinates, and at later times we introduce two new variables per vertex. Taking $n,q$ to denote the $q$-th vertex in increasing time order between sites $n$ and $n+1$, and following the labelling of Fig.~\ref{fig:vertex}, we define the energy transfer $\eta_{n,q}= \nu^\prime_{n+1} - \nu_{n+1}$ and the energy gain $\tilde{\eta}_{n,q} =(\nu^\prime_n + \nu^\prime_{n+1}) - (\nu_n + \nu_{n+1})$ at the vertex.
% by $\nu_n^\prime  = \nu_n -   \eta_{n,q} +\tilde{\eta}_{nq}$ and  $\nu_{n+1}^\prime = \nu_{n+1} + \tilde{\eta}_{n,q}$. 
 This transformation from $\nu^\prime_n$ and $\nu^\prime_{n+1}$ to $\eta_{n,q}$ and $\tilde{\eta}_{n,q}$ has unit Jacobian. Moreover, the right-hand side of Eq.~\eqref{eq:energyconservation} reduces to $2\pi \delta(\tilde{\eta}_{nq})$, allowing integration on each $\tilde{\eta}_{n,q}$. After making this transformation and integrating on $\tilde{\eta}_{n,q}$ at all $v$ vertices,  $v+L$ energy integration variables remain: the $L$ site energies at time zero and the $v$ energy transfers $\eta_{n,q}$.  

The integrand for the energy integrals contains two factors. One is a product of $2v+L$ functions $\rho(\nu_k)$ evaluated at the single-site energies $\nu_k$ appearing in the diagram. The other is a product of $2L$ $\delta$-functions on energies, from the definition of $P(\{\varepsilon_n \},\{ \varepsilon^\prime_n \};t)$ [Eq.~\eqref{eq:P}]. After the change of variables from $\{\varepsilon_n\}$ to $\{\omega_n\}$ and $\varepsilon_{\rm tot}$, introduced above Eq.~\eqref{eq:PtoM}, $L$ of these $\delta$-functions (the ones involving $\{\varepsilon_n  \}$) can be written as 
\begin{equation*}
    \delta(\varepsilon_{\rm tot} - \varepsilon^\prime_{\rm tot})\prod_{n=1}^{L-1} \delta(\omega_n - \sum_q \eta_{n,q})\,.
\end{equation*}
The remaining $L$ $\delta$-functions (the ones involving $\{\varepsilon^\prime_n\}$, which are associated with site propagators at time zero) fix the arguments of $L$ factors of $\rho(\nu_n)$ to be $\varepsilon^\prime_n$ for $1\leq n \leq L$. 
Summarising, these steps put the energy integral into the form 
\begin{multline}\label{eq:Mdiag}
    \delta(\varepsilon_{\rm tot} - \varepsilon^\prime_{\rm tot}) \prod_{n=1}^L \rho(\{\varepsilon^\prime_n\}) \int {\rm d} \{\eta_{n,q}\} \times \\
    \times\prod_{k=1}^{2v} \rho(\nu_k) \prod_{n=1}^{L-1} \delta(\omega_n - \sum_q \eta_{n,q}) \,.
\end{multline}
Here, integration runs over $v$ energy transfers $\{\eta_{n,q}\}$, and $\prod_{k=1}^{2v} \rho(\nu_k)$ consists of $2v$ factors with arguments $\nu_k$ that are linear combinations of $\{\varepsilon^\prime_n\}$ and $\{\eta_{n,q}\}$, determined by the structure of the diagram. Multiplying this by the result of the integral over mean time coordinates ((which is proportional to $t^v$)  and by the factor $(2\pi \lambda^2)^v$ [where $2\pi$ arises from Eq.~\eqref{eq:energyconservation} and $\lambda^2$ from the expansion in powers of $T_{n,n+1}$] gives a contribution from the diagram $\cal G$ that is of the form shown in Eq.~\eqref{eq:PDiagramContribution}.

This treatment of the energy integrals can be illustrated using the diagram shown in Fig.~\ref{fig:KtoP}(c). Here the $L+2v$ site energies are denoted $\varepsilon_n$, $\varepsilon_n^\prime$ and $\nu_n$ with $n=1$ and $n=2$. Integrating over relative time coordinates at both vertices and introducing energy transfers $\eta_1$ and $\eta_2$ for the right-hand and left-hand vertices respectively, we have  
\begin{eqnarray}
\nu_1 &=& \varepsilon_1^\prime - \eta_1, \qquad \nu_2 = \varepsilon_2^\prime + \eta_1, \nonumber \\ \varepsilon_1 &=& \varepsilon_1^\prime -\eta_1 - \eta_2, \qquad \varepsilon_2 = \varepsilon_2^\prime + \eta_1+\eta_2\,. \nonumber
\end{eqnarray}
Changing variables from $\varepsilon_1$ and $\varepsilon_2$ to $\omega_1 = \varepsilon^\prime_1 - \varepsilon_1$ and $\varepsilon_{\rm tot} = \varepsilon_1 + \varepsilon_2$, Eq.~\eqref{eq:Mdiag} for this example acquires the form
\begin{multline}\label{eq:Mdiag2}
    \delta(\varepsilon_{\rm tot} - \varepsilon^\prime_{\rm tot}) \prod_{n=1}^2 \rho(\{\varepsilon^\prime_n\}) \int {\rm d} \eta_1  \int {\rm d} \eta_2\, \times \\
    \times \rho(\varepsilon_1^\prime -\eta_1 )\rho(\varepsilon_2^\prime + \eta_1) \rho(\varepsilon_1^\prime - \omega_1) \rho(\varepsilon_2^\prime +\omega_1)\, \times \\ \times \delta(\omega_1 - \eta_1 - \eta_2)\,.
\end{multline}

As a further step, consider a diagram for $K(t)$ that similarly is fully connected and that contributes in the large $N$, small $\lambda$ limit. We again take an example without self-energy insertions and let $v$ denote the number of vertices. In this case evaluation involves integration over $2v$ site energies, and over the relative and mean time coordinates for each vertex. We now make use of two key points. The first point is that the integrand is invariant under a joint translation modulo $t$ of all the time coordinates for insertions of $T_{m,m+1}$ in the expansion of $e^{-i{\cal H}t}$ while keeping those from the expansion of $e^{i{\cal H}t}$ fixed (or alternatively with the roles of $e^{\pm i{\cal H}t}$ switched). Using a change of variables, we can therefore evaluate the relative time integrals by picking one vertex arbitrarily, fixing the relative time variable for that vertex to be zero, integrating over the $v-1$ remaining relative times, and multiplying the result by $t$. The second key point is that at small $\lambda$  the dominant contributions to integrals on relative times arise from a much smaller range than the dominant contributions to integrals on mean times. Specifically, these ranges are ${\cal O}(1)$ in the first case and ${\cal O}(\lambda^{-2})$ in the second case. With one relative time coordinate fixed to be zero, and other relative time coordinates all much smaller than the separation of mean time coordinates for vertices, each diagram for $K(t)$ is in a one-to-one correspondence with a diagram for $P(\{\varepsilon_n \},\{ \varepsilon^\prime_n \};t)$, as illustrated in Fig.~\ref{fig:KtoP}. To go from the former to the latter, the single-site propagators must be cut open at time $t$. This necessarily results in a leading-order diagram for $P(\{\varepsilon_n \},\{ \varepsilon^\prime_n \};t)$ if all relative time coordinates are zero, and it remains the case for dominant values of relative and mean time coordinates in the small $\lambda$ limit. To go between diagrams in the opposite direction, the external legs are closed: see again Fig.~\ref{fig:KtoP}.

Next we compare the contributions made to $K(t)$ and to $P(\{\varepsilon_n \},\{ \varepsilon^\prime_n \};t)$ by a pair of diagrams that are equivalent under this correspondence. To do so, we continue with the evaluation of a diagram for $K(t)$ with one relative time coordinate fixed to be zero, making a change of energy coordinates similar to the one introduced in the discussion of Eq.~\eqref{eq:Mdiag}. First, we use the variables $\{\varepsilon^\prime_n\}$ to label the $L$ sites energies at time zero. The site energies at time $t$ take the same values as at time zero because site propagators are closed in time [see Fig.~\ref{fig:KtoP}(c)], but it is convenient to treat them as independent variables $\nu_n^{\prime\prime}$ and to introduce the factor $\prod_{n=1}^L\delta(\nu_n^{\prime\prime} - \varepsilon^\prime_n)$ enforcing these constraints. Second, at each vertex we transform to the variables $\eta_{n,q}$ and $\tilde{\eta}_{n,q}$ as described above. After setting the relative time coordinate at one vertex to zero, $v-1$ integrals on relative time coordinates remain. They impose energy conservation at each of the $v-1$ vertices, so that all but one of the variables $\tilde{\eta}_{n,q}$ should be set to zero. At this stage we have $v+1$ energy integrals: over the single remaining $\tilde{\eta}_{n,q}$ and over all $v$ variables ${\eta}_{n,q}$. The integrand for this energy integral consists of $2v$ factors of $\rho(\nu_k)$, evaluated at the single-site energies $\nu_k$ appearing in the diagram, multiplied by $\prod_{n=1}^L\delta(\nu_n^{\prime\prime} - \varepsilon^\prime_n)$. The values of $\nu_n^{\prime\prime}$ can be expressed straightforwardly in terms of the integration variables. We take (without loss of generality) the vertex for which the relative time coordinate is fixed at zero to have the labels $(n,q) = (1,1)$. We then have $\nu_1^{\prime\prime} = \varepsilon^\prime_1 - \sum_q \eta_{1,q} + \tilde{\eta}_{1,1}$, $\nu_n^{\prime\prime} = \varepsilon^\prime_n - \sum_q \eta_{n,q} + \sum_q \eta_{n-1,q} $ for $1<n<L$ and $\nu_L^{\prime\prime} = \varepsilon^\prime_L + \sum_q \eta_{L-1,q}$. In consequence, the condition $\nu^{\prime\prime}_n = \varepsilon^\prime_n$ for all $n$ implies $\tilde{\eta}_{1,1}=0$ and $\omega_n=0$ for $n=1 \ldots L-1$. 

In summary, the form of the resulting integral is
\begin{multline}\label{eq:Kdiag}
    \int {\rm d}\{\varepsilon^\prime_n\} \int {\rm d} \{\eta_{n,q}\} \times \\
    \times \prod_{k=1}^{2v} \rho(\nu_k) \prod_{n=1}^{L-1} \delta\big(\sum_q \eta_{n,q}\big) \,.
\end{multline}
It is now apparent that for a pair of corresponding diagrams, the integrands in Eqns.~\eqref{eq:Mdiag} and \eqref{eq:Kdiag} are closely related. Specifically, the arguments $\nu_k$ in Eq.~\eqref{eq:Kdiag} are the same linear combinations of $\{\varepsilon^\prime_n\}$ and $\{\eta_{n,q}\}$ as those in Eq.~\eqref{eq:Mdiag}. Moreover, the integral on mean time coordinates is the same for both diagrams in the pair. These facts directly imply the relation given in Eq.~\eqref{eq:fully}, where the factor of $t$ is the one arising from joint translation of the relative time coordinates, and the factor of $(2\pi)^{-1}$ appears because evaluation of the diagram for $K(t)$ involves one fewer integral of the type displayed in Eq.~\eqref{eq:energyconservation} than evaluation of the equivalent diagram for $P(\{\varepsilon_n \},\{ \varepsilon^\prime_n \};t)$. The relation given in Eq.~\eqref{eq:fully} is the foundation for the main result of this paper, connecting $P(\{\varepsilon_n \},\{ \varepsilon^\prime_n \};t)$ and $K(t)$. To establish it in the necessary generality, we must also examine diagram with self-energy insertions, and ones that are not fully connected. 

\begin{figure}[tbh]
    \centering   \includegraphics[width=0.7\columnwidth]{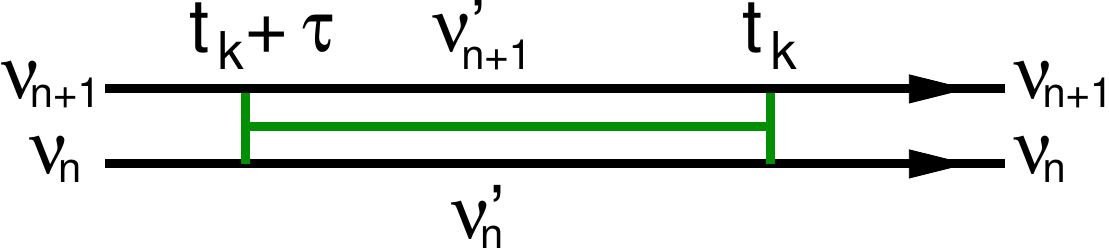}
    \caption{A self-energy insertion, labeled as in Eq.~\eqref{eq:selfenergy2}. 
}
    \label{fig:selfenergy}
\end{figure}

Diagrams with self-energy insertions retain the invariance discussed above, under joint translation of time coordinates, and so the equivalence we have described, between diagrams for $P(\{\varepsilon_n \},\{ \varepsilon^\prime_n \};t)$ and for $K(t)$, applies for these as well. Each self-energy insertion has associated with it mean and relative time coordinates, and two internal site energies. Integration over the internal site energies and the relative time coordinate give a result that is a function of the external site energies but not of the mean time coordinate. For the example illustrated in Fig.~\ref{fig:selfenergy} this calculation gives
\begin{multline}\label{eq:selfenergy2}
    \Sigma_{(\pm)}(\nu_n + \nu_{n+1}) = \int_0^\infty {\rm d} \tau  \int {\rm d}\nu^\prime_m\int {\rm d}\nu^\prime_{m+1}\,\times\\ \times 
      \rho(\nu^\prime_m)\, \rho(\nu^\prime_{m+1}) 
      e^{\pm i (\nu_m + \nu_{m+1} - \nu^\prime_m - \nu^\prime_{m+1})\tau}\,,
\end{multline}
where the sign is chosen according to the origin ($e^{\pm i{\cal H}t}$) of the self-energy insertion. The functions $\Sigma_{(\pm)}(\nu_n + \nu_{n+1})$ appear as factors in the energy integrands in a diagram for $P(\{\varepsilon_n \},\{ \varepsilon^\prime_n \};t)$ or for $K(t)$, generalising Eqns.~\eqref{eq:Mdiag} and \eqref{eq:Kdiag} respectively. Since the self-energy factors are the same in both cases, our main result [Eq.~\eqref{eq:fully}] holds for contributions from fully connected diagrams that include self-energy insertions.

\subsubsection{Treatment of diagrams that are not fully connected}\label{sec:proof2}

It remains to discuss diagrams that are not fully connected. For these diagrams, the vertices define a division of the system into subsystems, as discussed at the start of Sec.~\ref{sec:correspondence}. Consider evaluation of the contribution from such a diagram for $K(t)$, starting with integration over the relative time coordinate for each vertex. The integrand is invariant under joint translations modulo $t$ of all relative time coordinates, applied to each subsystem independently. We can therefore evaluate the relative time integrals by picking one vertex arbitrarily in each subsystem consisting of two or more sites, fixing the relative time variables for these vertices to be zero, and integrating over the remaining relative time coordinates. Fixing these relative time coordinates to be zero puts each diagram for $K(t)$ into a one-to-one correspondence with a diagram for $P(\{\varepsilon_n \},\{ \varepsilon^\prime_n \};t)$.

\begin{figure}[tbh]
    \centering   \includegraphics[width=0.9\columnwidth]{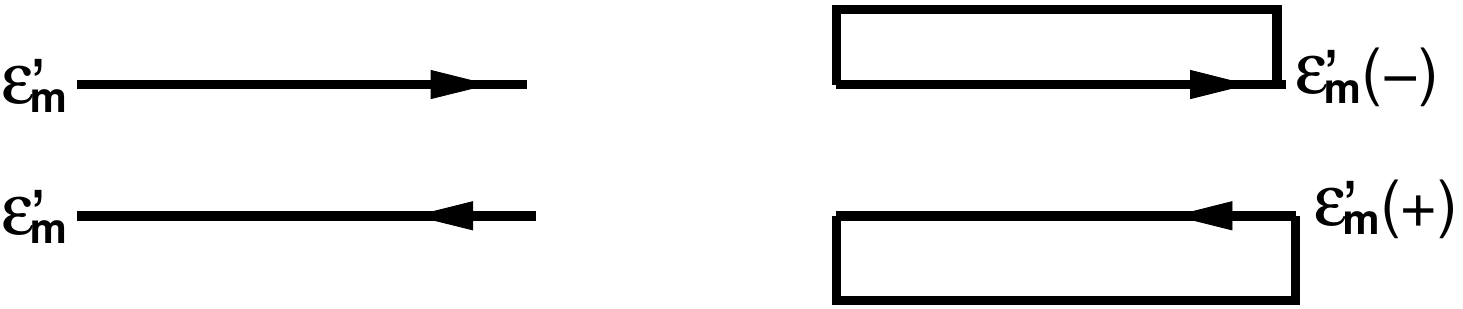}
    \caption{Illustration of the different contributions to $P(\{\varepsilon_n \},\{ \varepsilon^\prime_n \};t)$ and to $K(t)$ made by a subsystem consisting of a single site. See main text for details.
}
    \label{fig:singlesite}
\end{figure}

In this correspondence, single-site subsystems form a special case in two ways. First, integration over time coordinates for vertices does not involve single-site clusters, since by definition these sites are not linked by any vertices. Second, as illustrated in Fig.~\ref{fig:singlesite}, if no vertices involve the site $m$, two separate energy labels for this site appear in the diagram for $K(t)$, but only one in the diagram for $P(\{\varepsilon_n \},\{ \varepsilon^\prime_n \};t)$. Evaluation of the contribution to $K(t)$ involves summation over both these energy labels. The summand consists of the factor $e^{it[\varepsilon_m^\prime(+) - \varepsilon^\prime_m(-)]}$ multiplied in general by factors arising from self-energy insertions. If these involve the site $m$, they are functions either of $\varepsilon_m^\prime(+)$ or of $\varepsilon_m^\prime(-)$, rather than of a single variable $\varepsilon^\prime_m$, as is the case in the corresponding diagram for $P(\{\varepsilon_n \},\{ \varepsilon^\prime_n \};t)$. Unless further restrictions are imposed, this means that  contributions of a corresponding pair of diagrams which include single-site subsystems are not simply related. Simplifications arise, however, at early times and at late times, and for small $\lambda$ these time regimes overlap. 

At early times ($\lambda^2 t \ll 1$) the only relevant diagram is zeroth order in $\lambda$ and Eq.~\eqref{eq:SFFearly} follows straightforwardly. By contrast, after the onset of the ramp in the single-site SFF ($t\gg N^{1/2}$) the calculation simplifies because in that regime only the terms with $|\varepsilon^\prime_m(+) -\varepsilon^\prime_m(-)| \lesssim t^{-1}$ are important. We can re-write the ensemble-averaged double sum in terms of mean and relative energy coordinates, and use the mean energy coordinate $\varepsilon^\prime_m = [\varepsilon^\prime_m(+) +\varepsilon^\prime_m(-)]/2$ as the argument in factors arising from self-energy insertions involving site $m$. The relative energy coordinate then appears only in the factor 
\begin{equation}\label{eq:singlesite}
e^{it[\varepsilon_m^\prime(+) - \varepsilon^\prime  _m(-)]}
\end{equation}
and summing on it generates the function $k(\varepsilon^\prime_m,t)$, as discussed in Appendix~\ref{sec:crossover} [see Eq.~\eqref{eq:Def-k}]. At late times, the contribution to $K(t)$ from a diagram that contains single-site subsystems is therefore related to the corresponding contribution to $P(\{\varepsilon_m \},\{ \varepsilon^\prime_m \};t)$ in the same way as previously discussed for diagrams with only larger subsystems, but with a factor of $k(\varepsilon^\prime_m,t)$ for each single-site subsystem $m$, as shown in Eq.~\eqref{eq:fully}. With this, the derivation of our results for the SFF given in Eqns.~\eqref{eq:SFFearly}, \eqref{eq:SFFdecomp} and \eqref{eq:fully} is complete.

\subsubsection{Regimes of validity of expressions for $K(t)$}

It is useful to collect together the regimes for the validity of our results, and to examine their dependence on $N$ and $\lambda$. The relevant scales are set by energy exchange between sites, and by the onset time of the ramp in the single-site SFF, and there is overlap of the domains in which Eqns.~\eqref{eq:SFFearly} and \eqref{eq:SFFdecomp} apply if the time for energy exchange is much greater than the ramp onset time. 

To discuss this in detail, it is necessary to recall the form of the single-site SFF, Eq.~\eqref{eq:singlesite}. For $t\ll  N$, $K_1(t)$ is well approximated by the sum of two terms. One of these is the disconnected contribution $N^2 [J_1(2t)/t]^2$, which is the square of the Fourier transform of $\rho(\varepsilon)$ with $J_1(2t)$ denoting a Bessel function of the first kind. The other is the ramp term $2t/\pi$. The first dominates at early times and the second at late times, with crossover at the ramp onset time $t = t_{\rm dip}\sim N^{1/2}$. The value of $t_{\rm dip}$ grows with $N$ because the disconnected contribution decays as a power law at large $t$, reflecting the non-analytic behaviour of $\rho(\varepsilon)$ at the band edges. We can ensure that the time for energy exchange is much greater than the ramp onset time by taking $\lambda \ll N^{-1/4}$. If the single-site density of states were infinitely differentiable, the ramp onset time would be of order unity, and $\lambda \ll 1$ would be sufficient for overlap of the domains. 

A further simplification occurs if the single-site SFF can simply be approximated by a ramp after its onset time, without taking account of the plateau beyond the single-site Heisenberg time. This is possible if we take $\lambda$ in the range $N^{-1} \ll \lambda \ll N^{-1/4}$. To see why, note that the single-site Heisenberg time is $N$. If $\lambda^2N \gg 1$, then there are no contributions from the single-site SFF to the many-body SFF at or beyond the single-site Heisenberg time. The embodiment of the simplification is that for $N^{1/2} \ll t \ll N$  we find (see Appendix~\ref{sec:crossover}) $k(\varepsilon^\prime_m,t) = t/(2\pi)$, leading to the result shown in Eq.~\eqref{eq:SFFdecomp}. We choose to include the broader range $N^{-1} \ll \lambda \ll N^{-1/4}$ in our calculations because this is useful for comparison with numerical simulations.

\subsection{Translationally invariant model}\label{sec:TranslationInv}

A translationally invariant model in the same spirit as the one discussed above can be obtained by imposing periodic boundary conditions, and taking $H_n$ and $T_{n,n+1}$ in Eq.~\eqref{eq:Hamiltonian} to be independent of $n$, so that the factors of $\delta_{mn}$ are omitted from the right-hand sides of Eqns.~\eqref{eq:Hcorrelator} and \eqref{eq:Tcorrelator}. Here we outline the consequences of translational invariance without attempting a detailed analysis.

\begin{figure}[htb]
    \centering   \includegraphics[width=0.8\columnwidth]{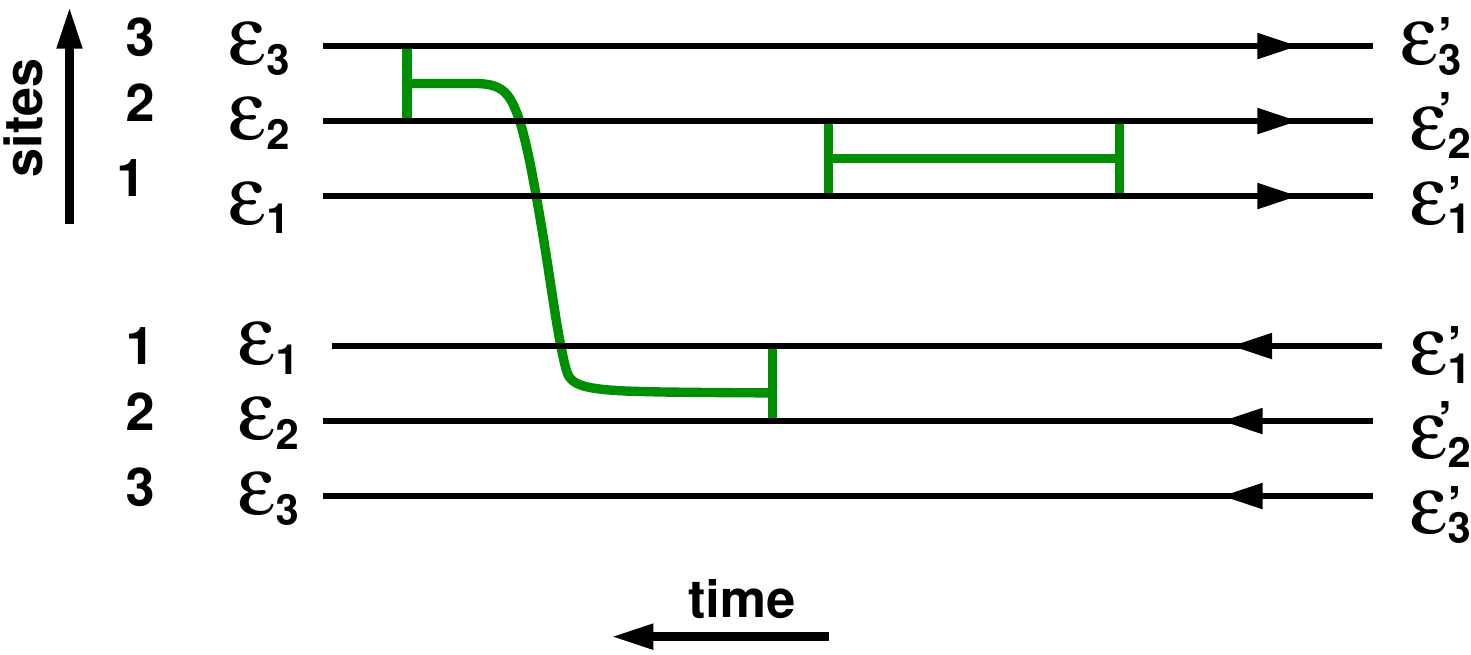}
    \caption{A contribution to $P(\{\varepsilon_k \},\{ \varepsilon^\prime_k \};t)$ that appears only in a translationally invariant model (see Fig.~\ref{fig:diagrams1} for notation and a comparison). The vertex in this diagram arises from the average $[T_{2,3}T_{1,2}]_{\rm av}$ which is zero in our model without translational invariance, but non-zero with translational invariance. 
}
    \label{fig:TranslationInv}
\end{figure}

New contributions to $P(\{\varepsilon_n \},\{ \varepsilon^\prime_n \};t)$ appear in the diagrammatic expansion for this translationally invariant model, as illustrated in Fig.~\ref{fig:TranslationInv}. These contributions are, however, suppressed by factors of $N^{-1}$. For the example shown in Fig.~\ref{fig:TranslationInv}, this suppression is because the vertex imposes $\varepsilon_1 = \varepsilon_3$ and $\varepsilon_1^\prime = \varepsilon_3^\prime$. Our results for energy dynamics at large $N$ and small $\lambda$ are therefore unchanged by the introduction of translational invariance in this form. 

By contrast, there are new contributions in the diagrammatic expansion for $K(t)$ for this translationally invariant model that are not suppressed at large $N$. In general, these additional terms appear difficult to analyse, but simple conclusions can be drawn for diagrams that are fully connected in the sense defined in Sec.~\ref{sec:correspondence}. These diagrams dominate at times $\lambda^2 t \gg (\ln L)^2$. For such fully-connected diagrams, there is a simple recipe that generates $L$ equal contributions to $K(t)$ in the translationally-invariant system from each contribution in the model we have studied without translational invariance.  These contributions are obtained by taking every vertex in the diagram and replacing it by a new one in which one end (say the one arising from the expansion of $e^{+i{\cal H}t}$) has been translated in space by $k \mod L$ lattice sites. All $L$ contributions are obtained from the values $k= 0,1 \ldots L-1$. Using as a convenient illustration diagrams for $P(\{\varepsilon_n \},\{ \varepsilon^\prime_n \};t)$ rather than for $K(t)$, Fig.~\ref{fig:TranslationInv} can be obtained from Fig.~\ref{fig:diagrams1} by a translation of this type with $k=2$ (note that $k=2$ is equivalent to $k=-1 \mod L$ for $L=3$).

In this way, we find for the late-time regime that $K(t)$ in the model with translational invariance is a factor of $L$ larger than without translational invariance. This conclusion is unsurprising, since translational invariance allows $\cal H$ to be block-diagonalised, with each of the $L$ blocks contributing additively to $K(t)$. Similar conclusions have been reached previously for translationally invariant random Floquet quantum circuits \cite{Chan_2022}.

\subsection{Two-site system}\label{sec:twosites}

In this section we study the two-site version of our model. First we solve the master equation to obtain the transition probability. From this we compute the two-point correlation function of energy density and the SFF. The two-site model is simple because in this case the transition probability depends only on a single coordinate $\omega_1$ (together with the conserved total energy $\varepsilon^\prime_{\rm tot}$), in contrast to the $L-1$ energy transfers $\{\omega_n\}$ in an $L$-site system

For $L=2$ the master equation [Eq.~\eqref{eq:master}] implies (dropping the subscript on $\omega_1$)
\begin{multline}\label{eq:twositemaster}
   \partial_t M(\omega,\{\varepsilon^\prime_n\},t) =\\ -\gamma(\varepsilon^\prime_{\rm tot}) \big[M(\omega,\{\varepsilon^\prime_n\},t)
   - M_{\infty}(\omega,\{\varepsilon^\prime_n\}) \big]
\end{multline}
with 
\begin{equation}\label{eq:gamma}
\gamma(\varepsilon^\prime_{\rm tot}) = 2\pi \lambda^2 \rho_{\rm tot}(\varepsilon_{\rm tot}^\prime)
\end{equation}
and
\begin{equation}
M_{\infty}(\omega,\{\varepsilon^\prime_n\}) =  \frac{ \rho(\varepsilon^\prime_1 - \omega) \rho(\varepsilon^\prime_2+\omega)}{\rho_{\rm tot}(\varepsilon_{\rm tot}^\prime)}\,.
\end{equation}

The solution is
\begin{multline}\label{eq:twositesoln}
    M(\omega,\{\varepsilon^\prime_n\},t) = \delta(\omega) e^{-\gamma(\varepsilon^\prime_{\rm tot}) t}\\ + M_{\infty}(\omega,\{\varepsilon^\prime_n\})\big[ 1- e^{-\gamma(\varepsilon^\prime_{\rm tot}) t}\big]\,.
\end{multline}
This solution exemplifies in a simple way the decomposition of the transition probability set out in Eq.~\eqref{eq:Minit}. Specifically, the first term on the right-hand side of Eq.~\eqref{eq:twositesoln} has the form $M_X(\{\varepsilon_n^\prime\},t)\delta(\omega)$ for the decomposition $X$ of the two-site system into two subsystems, each consisting of a single site; the second term has the form $M_X(\omega, \{\varepsilon_n^\prime\},t)$ for the decomposition $X$ in which the two sites are connected by energy exchange. Note also that the long-time limit of this solution matches the result for general $L$ given in Eq.~\eqref{eq:Mtinfty}.

Using this solution to the master equation and Eq.~\eqref{eq:energycorrelator2} we find the two-point correlator of energies
\begin{multline}\label{eq:predC11}
    C_{11}(t) = \int {\rm d}\varepsilon \int {\rm d}\eta \,\rho(\eta) \rho(\varepsilon-\eta)\, \eta^2 \, e^{-\gamma(\varepsilon)t}\\ + \int {\rm d}\varepsilon \, [1 - e^{-\gamma(\varepsilon)t}] B(\varepsilon)
\end{multline}
with 
\begin{equation}
B(\varepsilon) = \frac{\big[\int {\rm d}\eta\, \eta \,\rho(\eta) \rho(\varepsilon-\eta)\big]^2}{\rho_{\rm tot}(\varepsilon)}
\end{equation}
We also find $C_{11}(t) + C_{12}(t) =1$ for all $t$, reflecting energy conservation. We present a comparison of these analytical results with numerical simulations in Fig.~\ref{fig:NumericsSFF}, finding essentially perfect agreement. 

We calculate the SFF for the two-site system using the solution to the master equation and the expressions for general $L$ given in Sec.~\ref{sec:overview}. At early times ($\lambda^2 t \ll 1$) we simply have $K(t) = [K_1(t)]^2$ from Eq.~~\eqref{eq:SFFearly}. At times later than the onset of the ramp in the single-site SFF ($t\gg N^{1/2}$) we apply Eq.~\eqref{eq:SFFdecomp}. Here the sum on $X$ has two terms and we write $K(t) = K^{(1)}(t)+ K^{(2)}(t)$, where $K^{(1)}(t)$ and $K^{(2)}(t)$ originate respectively from the first and second terms on the right-hand side of Eq.~\eqref{eq:twositesoln}. This gives
\begin{equation}\label{eq:K^(1)}
    K^{(1)}(t) = \left( \frac{t}{2\pi}\right)^2 \int {\rm d}\varepsilon_1^\prime \int {\rm d}\varepsilon^\prime_2\, e^{-\gamma(\varepsilon^\prime_{\rm tot})t}
\end{equation}
and 
\begin{equation}\label{eq:K^(2)}
    K^{(2)}(t) = \frac{t}{2\pi} \int {\rm d}\varepsilon\, [1 - e^{-\gamma(\varepsilon)t}]\,.
\end{equation}
The result displayed in Eq.~\eqref{eq:K^(1)} holds for times shorter than the onset of the late-time plateau in the single-site SFF ($t\ll N$). For comparison with simulations, it is convenient to remove this restriction, which we do by using Eq.~\eqref{eq:fully} in place of Eq.~\eqref{eq:SFFdecomp}. This gives the generalised expression
\begin{equation}\label{eq:K^(1)generalised}
   K^{(1)}(t) =  \int {\rm d}\varepsilon \int {\rm d}\eta \, k(\varepsilon - \eta,t) k(\eta, t) e^{-\gamma(\varepsilon)t}
\end{equation}
where $k(\varepsilon,t)$ is defined in Eq.~\eqref{eq:Def-k}. Eq.~\eqref{eq:K^(1)generalised} reduces to Eq.~\eqref{eq:K^(1)} for $t\ll N$. In Fig.~\ref{fig:NumericsSFF} we present a comparison of the analytical results for the SFF contained in Eqns.~\eqref{eq:K^(2)} and \eqref{eq:K^(1)generalised}  for two different  values of $\lambda$, again with essentially perfect agreement. 

\section{Master equation}\label{sec:mastersimulations}

Next we study various aspects of the master equation [Eq.~\eqref{eq:master}] for energy dynamics in a system of many sites. In Sec.~\ref{sec:mastergeneral} we set out some general properties of this equation. Then in Sec.~\ref{sec:masternumerics} we present results from numerical simulations, demonstrating that energy dynamics is diffusive. In Sec.~\ref{sec:masteranalytical} we calculate an approximate value for the energy diffusion constant, and in Sec.~\ref{sec:finiteT} we examine the temperature dependence of energy dynamics. In Sec.~\ref{sec:masternoisydiffusion} we approximate the master equation using a noisy diffusion equation and employ this to determine the long-time behaviour of the many-body return probability, showing that it is controlled by the timescale for energy diffusion across the system. Finally, in Sec.~\ref{sec:masterlate} we study the probability that there has been no energy transferred between two neighbouring sites as a function of time, analysing its approach to zero with increasing time; this controls the SFF at intermediate times, as indicated in Eq.~\eqref{eq:Kw}.

\subsection{General properties of the master equation}\label{sec:mastergeneral}

It is straightforward to show by direct substitution that the solution to the master equation obeys
\begin{equation}
    \partial_t \int {\rm d}\{\varepsilon_n\} \, P(\{\varepsilon_n \},\{ \varepsilon^\prime_n \};t) =0\,.
\end{equation}
This confirms that the transition rate $W(\{\nu_n\},\{ \varepsilon_n\})$ derived from the diagrammatic analysis of Sec.~\ref{sec:calculations} does indeed lead to a classical master equation. 

In a similar way
\begin{equation}
    \partial_t \int {\rm d}\{\varepsilon_n\} \,\sum_n \varepsilon_n\, P(\{\varepsilon_n \},\{ \varepsilon^\prime_n \};t) =0\,,
\end{equation}
showing that total energy is conserved in the process represented by this master equation. 

One can also check that the long-time form for $P(\{\varepsilon_n \},\{ \varepsilon^\prime_n \};t)$, obtained by substituting Eq.~\eqref{eq:Mtinfty} into Eq.~\eqref{eq:PtoM}, is a time-independent solution to the master equation, in which no net energy current flows. 

\subsection{Simulations}\label{sec:masternumerics}

To do numerical simulations, we generate stochastic trajectories for the master equation [Eq.~\eqref{eq:master}] using the Gillespie algorithm~\cite{gillespie1977exact,gillespie2007stochastic}. Since Eq.~\eqref{eq:master} is homogeneous in $\lambda^2$, the dynamics can be simulated in terms of the rescaled, dimensionless time $\tau=\lambda^2 t$.

Each site is initialized in a random configuration, with energies drawn from the single-site density of states $\rho(\varepsilon)$ given in Eq.~\eqref{eq:rho}.  
At each step of the simulation, one bond $m$ between sites $m$ and $m+1$ is chosen with probability
\begin{align}
p_m(\{\varepsilon_n\})=\frac{\gamma(\varepsilon_{m}+\varepsilon_{m+1})}{\sum_n \gamma(\varepsilon_{n}+\varepsilon_{n+1})},
\end{align}
where $\gamma(\varepsilon_{n}+\varepsilon_{n+1})$ is defined for a pair of sites in Eq.~\eqref{eq:gamma}.

The waiting time $\Delta\tau$ until the next event is then sampled from the exponential distribution
\begin{multline}
    P(\Delta \tau)=\Biggl(\sum_n \gamma(\varepsilon_{n}+\varepsilon_{n+1})\Biggr)\times 
\\  \times \exp\!\Biggl[-\Delta\tau \sum_n \gamma(\varepsilon_{n}+\varepsilon_{n+1})\Biggr].
\end{multline}
Once bond $m$ is selected, the pair of site energies is updated according to
\begin{align}
(\varepsilon_m,\varepsilon_{m+1}) \;\longrightarrow\; (\varepsilon_m',\varepsilon_m+\varepsilon_{m+1}-\varepsilon_{m}'),
\end{align}
with transition probability density
\begin{align}
\begin{split}
p\bigl((\varepsilon_m,\varepsilon_{m+1})\to(\varepsilon_m',\varepsilon_m+\varepsilon_{m+1}-\varepsilon_{m}')\bigr)
\\=
\frac{2\pi\lambda^2\,\rho(\varepsilon_m')\,\rho(\varepsilon_m+\varepsilon_{m+1}-\varepsilon_{m}')}{\gamma(\varepsilon_{m}+\varepsilon_{m+1})}.
\end{split}
\end{align}

In summary, each update consists of
\begin{align}
\begin{split}
\tau &\;\longrightarrow\; \tau+\Delta\tau, \\
(\varepsilon_m,\varepsilon_{m+1}) &\;\longrightarrow\; (\varepsilon_m',\varepsilon_m+\varepsilon_{m+1}-\varepsilon_{m}').
\end{split}
\end{align}
All results are obtained for chains with system size $L=128$ and open boundary conditions, averaged over $10^{10}$ trajectories.

\begin{figure*}[t!]
    \centering   \includegraphics{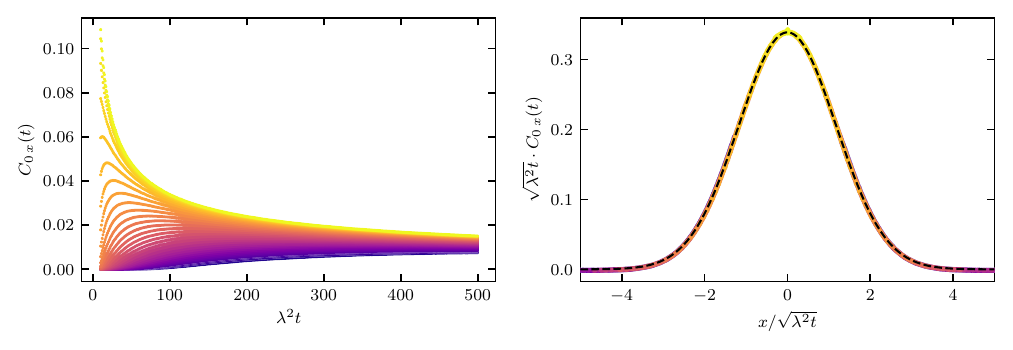}
    \caption{Behaviour of the two-point correlator of energy density. (a) $C_{0\,x}(t)$ obtained from the master equation Eq.~\eqref{eq:master} with $L=128$, averaged over $10^{10}$ realizations, for $t<10<500$, and $|x|<L/4$. Lighter colour denotes a smaller $|x|$. (b) Collapse of $\sqrt{t}C_{0\,x}(t)$ vs ${x}/{\sqrt{t}}$, and comparison with diffusive form [Eq.~\eqref{eq:scalingcollapse}] taking $D=0.692 \lambda^2$ (dashed line). The collapse shows perfect agreement with the diffusive prediction. 
}
    \label{fig:Diffusioncollapse}
\end{figure*}

To investigate the energy transport implied by Eq.~\eqref{eq:master}, we compute the autocorrelator 
\begin{align}\label{eq:Autocorrelator}
    C_{0\,x}(t)=\braket{\varepsilon_{0}\varepsilon_{x}(t)}
\end{align}
taking the origin for the position coordinate to lie at the centre of the system.
If the energy transport is described by diffusive behavior, $ C_{0\,x}(t)$ admits at large $x$ and $t$ the scaling collapse
\begin{align}\label{eq:scalingcollapse}
    \lim_{x\rightarrow\infty} \lim_{t\rightarrow\infty}C_{0\,x}(t)=\frac{1}{\sqrt{4\pi D t}}e^{-\frac{x^2}{4 D t}}.
\end{align}
The results for $C_{0\,x}(t)$ and the scaling collapse are shown in Fig.~\ref{fig:Diffusioncollapse}. They demonstrate perfect agreement with diffusive scaling and yield $D=0.692 \lambda^2$.

\subsection{Value of the energy diffusion constant}\label{sec:masteranalytical}

We can obtain an approximate value for the energy diffusion constant in the model by calculating the current generated by small amplitude modulations in the energy density. First, consider the probability distribution of the energy $\varepsilon$ at one site in a large system. It is given by
%Let $\varepsilon_{\rm av}$ be the energy density, so that $\varepsilon_{\rm tot} = L\varepsilon_{av}$. This probability distribution is 
\begin{equation}\label{eq:Ansatzzerothorder}
Z^{-1} \rho(\varepsilon) \,e^{-\beta \varepsilon} \quad \mbox{with} \quad Z = \int {\rm d}\varepsilon \,\rho(\varepsilon) e^{-\beta \varepsilon}\,.
\end{equation}
%\begin{equation}
 %    \rho(\varepsilon) \rho_{\rm tot,\,L-1}(\varepsilon_{\rm av}\,L - \varepsilon) 
%\end{equation}
Here $\beta$ is a Lagrange multiplier with the obvious interpretation as inverse temperature, which has a value fixed by the total energy in the system. For small $\beta$ one has at leading order $\langle \varepsilon \rangle = -\beta$, using $\int {\rm d}\varepsilon \, \varepsilon^2 \rho(\varepsilon) = 1$.

Now suppose a state is initiated with an energy distribution parameterised by small values of $\beta_n$ that vary with the site index $n$. 
We can calculate the current in this state using results given in Sec.~\ref{sec:twosites} for a pair of sites. Consider sites $n$ and $n+1$ with initial energies $\varepsilon^\prime_n$, $\varepsilon^\prime_{n+1}$, and energies $\varepsilon^\prime_n - \omega$, $\varepsilon^\prime_{n+1} +\omega$ at time $t$. The energy transfer from site $n$ to site $n+1$ is $\omega$. From Eq.~\eqref{eq:twositesoln} the current, given the initial energies, is
\begin{eqnarray}
J_{n\to n+1} &=& \partial_t \left. \langle \omega \rangle \right|_{t=0}\nonumber \\ &=& (\varepsilon^\prime_n - \varepsilon_{n+1}^\prime) \times \pi \lambda^2 \rho_{\rm tot}(\varepsilon_{\rm tot}^\prime)\,,
\end{eqnarray}
where $\varepsilon^\prime_{\rm tot} = \varepsilon_n^\prime + \varepsilon_{n+1}^\prime$ and the total density of states $\rho_{\rm tot}(\varepsilon^\prime_{\rm tot})$ is for the pair of sites. Averaging this current on $\varepsilon^\prime_n$ and $\varepsilon^\prime_{n+1}$ in the state described gives
\begin{equation}
    \langle J_{n\to n+1} \rangle = D(\beta_{n+1} - \beta_n) 
\end{equation}
with 
\begin{multline}\label{eq:Dapprox}
D = \frac{\pi \lambda^2}{2} \int {\rm d}\varepsilon^\prime_n \int {\rm d} \varepsilon^\prime_{n+1} (\varepsilon^\prime_n - \varepsilon^\prime_{n+1})^2 \times \\ \times\rho(\varepsilon^\prime_n) \rho(\varepsilon^\prime_{n+1}) \rho_{\rm tot}(\varepsilon_{\rm tot})\,.
\end{multline}
Using the relation between $\beta_n$ and $\langle \varepsilon_n\rangle$ and the continuity equation, we have
\begin{equation}
    \partial_t\langle \varepsilon_n\rangle = D [\langle \varepsilon_{n-1} \rangle + \langle \varepsilon_{n+1} \rangle - 2 \langle \varepsilon_{n} \rangle]\,,
\end{equation}
making it clear that $D$ should be interpreted as the energy diffusion constant. Evaluating Eq.~\eqref{eq:Dapprox} numerically gives the value $D =  0.7022\lambda^2$, which is close to but distinct from the value derived from Fig.~\ref{fig:Diffusioncollapse}, and from our most precise result, $D = 0.691(1)\lambda^2$ obtained from simulations of the master equation, as described in Sec.~\ref{sec:masternumerics}.
We show in Appendix~\ref{sec:Diffusionimprovedestimate} that the discrepancy between analytical and numerical estimates for the diffusion constant can be reduced by systematically extending the ansatz Eq.~\eqref{eq:Ansatzzerothorder}.

\subsection{Temperature dependence of dynamics}\label{sec:finiteT}

The previous calculation can be generalized to obtain estimates for the diffusion constant at finite inverse temperature $\beta$.
To do so, the probability distribution at site $n$ is characterized by a probability distribution
\begin{equation}\label{eq:Ansatzzerothorder2}
Z_{\beta+\beta_n}^{-1} \rho(\varepsilon) \,e^{-(\beta+\beta_n) \varepsilon} \quad
\end{equation}
 with
 \begin{equation}
  Z_{\beta+\beta_n} = \int {\rm d}\varepsilon \rho(\varepsilon
)\,e^{-(\beta+\beta_n) \varepsilon}\,.
\end{equation}
Here a potential difficulty presents itself, since $Z_{\beta+\beta_n}$ is in general realisation dependent, necessitating the use of replica or other techniques to handle the disorder average. However, relative fluctuations in $Z_{\beta+ \beta_n}$ are subleading in $N^{-1}$ and $\lambda$. As a consequence, for large $N$ and small $\lambda$ limit we can simply use the disorder-averaged value of $Z_{\beta+\beta_n}$ where it appears in the following calculations.

Using the notation
\begin{align}
\braket{\varepsilon}_\beta=Z_{\beta}^{-1} \int {\rm d}\varepsilon\, \varepsilon \rho(\varepsilon) \,e^{-\beta \varepsilon}
\end{align}
and analogously for $\braket{\varepsilon^2}_\beta$, the expectation value for the energy at leading order is
    \begin{align}\label{eq:energybeta}
	\begin{split}
		\langle \varepsilon_n\rangle &=
        %\braket{\varepsilon_n}_\beta-\beta_n(\braket{\varepsilon^2}_\beta-\braket{\varepsilon}_\beta^2)=
        \braket{\varepsilon_n}_\beta-\beta_n \chi_\beta^2 
	\end{split}
	\end{align}
with
\begin{equation}
    \chi_\beta^2 = \braket{\varepsilon^2}_\beta-\braket{\varepsilon}_\beta^2\,.
\end{equation}

Inserting the relation Eq.~\eqref{eq:energybeta} between Lagrange multiplier $\beta_n$ and energy density $\braket{\varepsilon_n}$ into the time evolution of $\langle \varepsilon_n \rangle -\braket{\varepsilon}_\beta$ leads to a diffusion equation with temperature-dependent diffusion constant
\begin{align}\label{eq:Dbeta}
\begin{split}
    D_\beta=\frac{2 \pi \lambda^2}{Z_\beta^2 \chi_\beta^2}
\int {\rm d}\varepsilon_1\,{\rm d}\varepsilon_2\,{\rm d}\varepsilon_3\,
\rho(\varepsilon_1)\rho(\varepsilon_2)\rho(\varepsilon_3)
\times \\\times \rho(\varepsilon_1+\varepsilon_2-\varepsilon_3)
e^{-\beta(\varepsilon_1+\varepsilon_2)}
\left(\varepsilon_1^2-\varepsilon_1\varepsilon_3\right)\,,
\end{split}
\end{align}
which reduces to Eq.~\eqref{eq:Dapprox} for $\beta=0$. 

This expression is in excellent agreement with the numerical results~(see Appendix~\ref{sec:Diffusionimprovedestimate}), with an accuracy that increases as $\beta$ increases. In the low temperature limit $\beta\rightarrow \infty$, the leading order expression can be obtained using an expansion of $\rho(\varepsilon)$ around the band edge, giving
\begin{align}\label{eq:Dbeta1}
    D_{\beta}=\frac{15\lambda^2}{4\beta^2}\,.
\end{align}
We note that the low-temperature scaling of the diffusion constant is not universal. Instead, it depends on the specific single-site density of states.

\subsection{Noisy diffusion equation}\label{sec:masternoisydiffusion}

We have shown in Sec.~\ref{sec:overview} [see Eq.~\eqref{eq:SFFdecomp}] that the approach of the SFF to a ramp is controlled by the return probability for the many-body energy distribution. In addition, we have shown in Sec.~\ref{sec:masternumerics} using numerical simulations that the dynamics of the energy distribution is diffusive at long times. The simulations, however, do not give direct access to the return probability because the algorithm used generates energy trajectories in time; the probability density for these to return to the origin decreases exponentially with system size, and so is too small to measure in large systems. For this reason and in order to provide insight more generally, we outline in this section an approximate analytical treatment of energy dynamics. The approximation consists of replacing the master equation with a noisy diffusion equation. This is expected to represent the universality class for the dynamics, because the non-linear contributions that are neglected stem from the density-dependence of the diffusion coefficient and these are irrelevant in the renormalisation group sense \cite{Spohn_2016}.

Our starting point is a phenomenological evolution equation for energies $\varepsilon_n(t)$ at sites $n$ as a function of time $t$, in the form
\begin{multline}\label{eq:noisydiffusion}
    \partial_t \varepsilon_n(t) =D[\varepsilon_{n+1}(t)+ \varepsilon_{n-1}(t) - 2\varepsilon_n(t)]\\ + \omega_{n-1}(t)- \omega_n(t)\,.
\end{multline}
Here $D$ is a parameter with an obvious interpretation as the energy diffusion constant, and $\omega_m(t)$ is white noise with mean zero, strength $\mu$ and correlator
\begin{equation}
    \langle \omega_m(t_1)\omega_n(t_2) \rangle_\omega = \mu\,\delta_{mn}\,\delta(t_1 - t_2)\,,
\end{equation}
where $\langle \ldots \rangle_\omega$ indicates an average over noise histories. This evolution equation is much simpler to study than the master equation because it is linear in the energies $\varepsilon_n(t)$, whereas the transition rates in  Eq.~\eqref{eq:master} are non-linear in these variables. The first (diffusive) term in Eq.~\eqref{eq:noisydiffusion} drives the energy distribution towards to a uniform one, and is intended to reflect the fact that the full, non-linear transition rate does the same on average. The second (noise) term in Eq.~\eqref{eq:noisydiffusion} introduces fluctuations in the evolution, which are also present in the master equation. 

The evolution equation can be solved straightforwardly by Fourier transform and integration. Consider a system of $L$ sites. For convenience we impose periodic boundary conditions and take $L$ odd, writing $L=2K_{\rm max}+1$ with $K_{\rm max}$ integer. Let 
 \begin{equation}
     \tilde{\varepsilon}_k(t) = L^{-1/2}\sum_n e^{-ikn}\varepsilon_n(t)
 \end{equation}
 with $k = 2\pi K/L$ and $K$ integer, lying in the range $[-K_{\rm max},\, K_{\rm max}]$. By design, the mode $\tilde{\varepsilon}_k(t)$ with $k=0$ has no dynamics, while the other modes relax at a $k$-dependent rate to an equilibrium distribution that is Gaussian.
 Averaging over noise histories and over initial values for $\tilde{\varepsilon}_k(t)$ drawn from this equilibrium distribution, we find
 \begin{equation}
     \langle [\tilde{\varepsilon}_k(t)- \tilde{\varepsilon}_k(0) ][\tilde{\varepsilon}_q(t)- \tilde{\varepsilon}_q(0) ]\rangle_\omega \\
= \delta_{k,-q}\frac{\mu}{D}  \varphi_k(t)
 \end{equation}
with
\begin{equation}
    \varphi_k(t) = 1- e^{-2D(1-\cos k)t}\,.
\end{equation}
Moreover, since $\tilde{\varepsilon}_k(t)- \tilde{\varepsilon}_k(0)$ has a Gaussian distribution, the return probability for this complex coordinate is inversely proportional to its variance. Combining contributions from all modes, the many-body return probability is therefore enhanced relative to its long-time value by the factor $\prod_{K=1}^{K_{\rm max}}  \left[ \varphi_k(t)\right]^{-1}$. For $Dt\gg1$ we have 
\begin{equation}
     \prod_{K=1}^{K_{\rm max}}  \left[ \varphi_k(t)\right]^{-1} - 1 \sim \sum_{K=1}^\infty \exp(-4\pi^2 K^2 Dt/L^2)\,.
\end{equation}
This asymptotic behaviour makes it apparent that the timescale for approach to the limiting value of the many-body return probability is $t_{\rm Th} = L^2/D$. Similar expressions have been obtained previously for Floquet circuits with a conserved density \cite{Friedman_2019}, and from a hydrodynamic theory of the SFF \cite{Winer_2022}. 

\subsection{Energy transfers: correlations and late-time behaviour}\label{sec:masterlate}

In this section we discuss the behaviour of the quantity $w(t)$, introduced in Sec.~\ref{sec:manysites}, which gives the probability for there to have been no energy exchange between a specified neighbouring pair of sites before time $t$. It has initial value $w(0)=1$ and decreases monotonically towards zero at late times. We require its form in order to draw conclusions about the behaviour of the SFF at intermediate times, via Eq.~\eqref{eq:Kw}. We first present evidence that energy exchange is almost uncorrelated between different pairs of neighbouring sites. Then we show that late-time decay of $w(t)$ is sub-exponential.

\subsubsection{Correlations}\label{sec:correlations}

\begin{figure*}[t!]
    \centering   
    \includegraphics{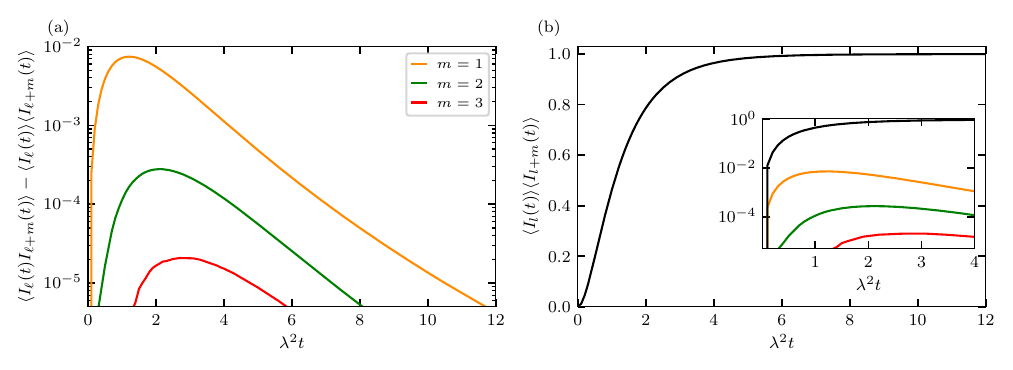}
        \caption{ 
    Correlations in energy transfers between nearby sites. (a) The connected correlator $\langle I_\ell(X) I_{\ell +m}(X)\rangle - \langle I_\ell(X)\rangle \langle I_{\ell +m}(X)\rangle$ for different distances $m$, obtained by averaging over $10^{10}$ realizations for $L=128$ and binning over time steps $\delta t=0.1$. (b) $\langle I_\ell(X)\rangle \langle I_{\ell +1}(X)\rangle$ as a function of time. Inset: The data $\langle I_\ell(X)\rangle \langle I_{\ell +1}(X)\rangle$ compared to $\langle I_\ell(X) I_{\ell +m}(X)\rangle - \langle I_\ell(X)\rangle \langle I_{\ell +m}(X)\rangle$ on the same scale. The connected correlator is more than one order of magnitude smaller than the disconnected component for any $m\geq 1$.}
    \label{fig:transfercorrelns}
\end{figure*}

It is useful to use the indicator function $I_\ell(X)$, introduced above Eq.~\eqref{eq:Minit}. Then $w(t) = 1 - \langle I_\ell(X)\rangle$, where the average is over decompositions $X$ occurring at time $t$. Correlations in energy transfer between neighbouring sites are characterised by $\langle I_\ell(X) I_{\ell +m}(X)\rangle - \langle I_\ell(X)\rangle \langle I_{\ell +m}(X)\rangle$. Simulation results for this quantity are shown in Fig.~\ref{fig:transfercorrelns}. Correlations are small for $m=1$ and decrease rapidly with increasing $m$.

\subsubsection{Late-time behaviour}

We find that the decay of $w(t)$ with $t$ is initially exponential, with a crossover to a stretched exponential at late times. The slower decay at late times is associated with initial energy distributions $\{\varepsilon^\prime_n\}$ in which the energies for a sequence of sites $m$ lie close to the upper or lower band edge of the single-site density of states $\rho(\varepsilon^\prime_m)$. 

A  related mechanism leading to sub-exponential decay in quantum systems with finite local Hilbert space dimension was identified in Ref.~\cite{mcculloch2025subexponentialdecaylocalcorrelations}. The similarity between this mechanism and the one that we discuss is that both arise from rare initial states, involving a void in the particle distribution in \cite{mcculloch2025subexponentialdecaylocalcorrelations}, and a region with extremal energy density in our case. One important difference is that that the discussion of Ref.~\cite{mcculloch2025subexponentialdecaylocalcorrelations} is specific to a quantum system, whereas in our case, although the underlying problem we treat is quantum mechanical, stretched exponential decay arises within the framework of a classical master equation. A second difference is that the void in particle number of Ref.~\cite{mcculloch2025subexponentialdecaylocalcorrelations} involves a discrete variable, while energy density in our case is a continuous variable.

Our starting point for calculations is the expression for the rate $\gamma(\varepsilon^\prime_{\rm tot})$ of energy exchange, given in Eq.~\eqref{eq:gamma}. Whilst this quantity has been introduced in the context of the two-site model, it applies equally to a pair of neighbouring sites that are part of a larger system, provided only that the total energy accommodated on the pair of sites is $\varepsilon^\prime_{\rm tot}$. If this total energy is independent of time, as it is for a two-site system, then $w(t) = e^{-\gamma(\varepsilon^\prime_{\rm tot})t}$. Conversely, if the pair of sites are part of a larger system, then this probability should be appropriately averaged over histories for $\varepsilon^\prime_{\rm tot}$, which in turn depend on the initial energy distribution. We are not able to carry out this average exactly; instead we discuss two approximations, which are appropriate at early and late times, respectively.

At early times, we average the decay rate on energy, writing 
\begin{equation}
\gamma = \int {\rm d}\varepsilon_{\rm tot}^\prime \, \rho_{\rm tot}(\varepsilon^\prime_{\rm tot})\, \gamma(\varepsilon^\prime_{\rm tot}) = 1.214  \lambda^2\,,
\end{equation}
where $\rho_{\rm tot}(\varepsilon_{\rm tot}^\prime)$ is the total density of states for the pair of sites under consideration and the right-hand side of this equation is obtained by evaluating the integral numerically. Then
\begin{equation}\label{eq:expdecay}
w(t) \approx e^{-\gamma t}\,.
\end{equation}
This exactly captures the initial behaviour. 

To treat late times, we build on the fact that $\gamma(\varepsilon^\prime_{\rm tot})$ falls to zero as $\varepsilon^\prime_{\rm tot}$ approaches the band edges. For definiteness, we focus on the lower band edge at $\varepsilon_m = -2$ and expand around this energy, writing $u_m = \varepsilon_m +2$ and $u = \varepsilon_{\rm tot} +4$ . Then at leading order in $u_m$ and $u$ we find $\rho(\varepsilon_m) = \sqrt{u_m}/\pi$ and $\gamma(\varepsilon^\prime_{\rm tot}) = u^2_{\rm}\lambda^2$. We expect that the dominant contribution to $w(t)$ at large $t$ will arise from initial energy distributions in which, first, $u_{\rm}$ is small at $t=0$ for the specified pair of sites, and second, initial energies are small, with $u_m \sim u$ inside an energy void surrounding these sites, so that $u_{\rm}$ for the specified pair of sites remains small until time $t$. Since energy diffuses into this void from more distant parts of the system, the minimum size required for the void is $L_{\rm void} \sim \sqrt{Dt}$, with $D$ the energy diffusion constant $D$. 

The probability density for such a void to occur in the average over initial site energies $\{ \varepsilon_n^\prime\}$ is given by the total density of states $\rho_{\rm tot}(\varepsilon_{\rm tot})$ evaluated from Eq.~\eqref{eq:rhotot} for a system of length $L_{\rm void}$ with total energy $\varepsilon_{\rm tot} = L_{\rm void}(u-2)$. For $L_{\rm void}u \ll 1$ this is determined by the form of the density of states near the band edge. For an $L$-site system we have from Eq.~\eqref{eq:rhotot} 
\begin{eqnarray}
\rho_{\rm tot}(L[u-2]) &\sim& \pi^{-L} \int_0^\infty {\rm d}u_1 \ldots \int_0^\infty {\rm d}u_L \nonumber \\ &&(u_1 \ldots u_L)^{1/2} \delta(Lu - \sum_k u_k)\nonumber \\ && \propto u^{3L/2-1}\,.
\end{eqnarray}
Hence the total density of states in a system of length $L_{\rm void}$ is proportional to $u^{\kappa L_{\rm void} -1}$ with $\kappa =3/2$. By this route, our estimate for the late-time behaviour of $w(t)$ is
\begin{equation}
    w(t) \propto  \int_0^\infty {\rm d} u\, u^{\kappa \sqrt{Dt}-1} e^{-u^2 \lambda^2 t}\,.
\end{equation}
Evaluating this integral for large $t$ using the saddle-point approximation, we find the asymptotic behaviour
\begin{equation}\label{eq:wlate}
    w(t) = A \exp(-\frac{\kappa}{4}\sqrt{Dt} \ln \lambda^2 t ) \approx Ae^{-b\sqrt{\lambda^2t}}
\end{equation}
where $A$ and $b$ are constants, and in the right-most expression we have simplified the argument of the exponential by omitting the logarithm.
    \begin{figure}[t!]
    \centering   
    \includegraphics{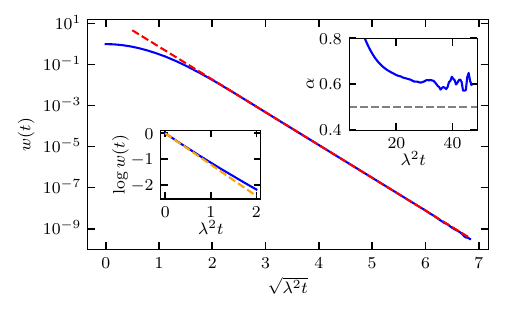}
    \caption{ 
    Behaviour of $w(t)$. Data (blue) from simulations of the master equation, and fit (red) to $Ae^{-b\sqrt{\lambda^2t}}$ [Eq.~\eqref{eq:wlate}]. Upper right inset: dependence on $t$ of the effective exponent $\alpha \equiv {\rm d}\ln[- \ln w(t)] / {\rm d}\ln t$ (dashed line: $\alpha = 1/2$). Lower left inset: early time behaviour, compared with Eq.~\eqref{eq:expdecay}.
}
    \label{fig:Numericswt}
\end{figure}

Numerical results for \(w(t)\) are shown in Fig.~\ref{fig:Numericswt}. The late-time data are compatible with a stretched-exponential decay; however, the numerical resolution is insufficient to determine conclusively whether the stretched exponent is one half. Early-time data match Eq.~\eqref{eq:expdecay} well.
    
\section{SFF in a model with $N=2$}\label{sec:N=2}

\begin{figure}[t!]
    \centering   \includegraphics{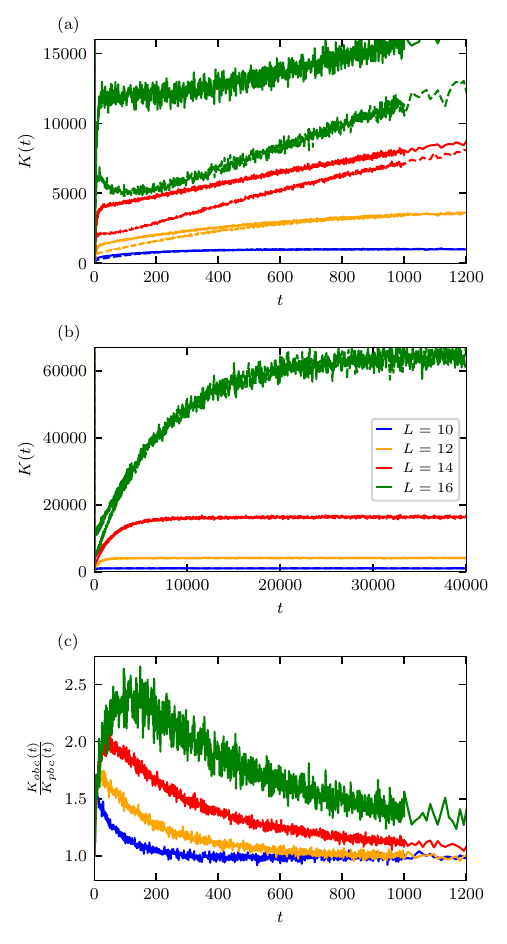}
    \caption{$K(t)$ for the Hamiltonian of Eq.~\eqref{eq:Hamlocal} with open boundary conditions~(solid) and periodic boundary conditions~(dashed) and different system sizes: (a) at early times, and (b) at times extending beyond the Heisenberg time. (c) Ratio between the spectral form factor for open and periodic boundary conditions. 
    All results are averaged over 5000 realizations~(1000 for $L=16$), and system sizes are as indicated in (b).
}
    \label{fig:Localmodel}
\end{figure}

An obvious and important question is whether the results we have obtained (in particular, the existence of a peak in the SFF at intermediate times) apply generally or are specific to models like the one we have studied, that have large local Hilbert space dimension and weak coupling between sites. One might at first suspect that models with small $N$ and strong intersite coupling behave differently, since as far as we know, no attention has been drawn previously to an intermediate-time peak in the SFF in such models. In fact, as we show in this section, the same qualitative features are present in the ergodic phase of a standard model for many-body localisation, which has $N=2$ and strong intersite coupling.

As discussed in Sec.~\ref{sec:SFFIIA}, enhancement of the SFF arises from decomposition of the system into two or more subsystems between which there has been no energy exchange. This is a more prominent effect in systems with open boundary conditions than with periodic boundary conditions, since two subsystems are separated by a single interface in the first case, and by two interfaces in the second case, and each interface is weighted by the factor $w(t)$, which is small at late times. The effect also becomes more prominent with increasing system size, since the number of ways to divide a system into multiple subsystems grows with size. 

To test these predictions in a model with a small local Hilbert space dimension, we consider the spin-half Hamiltonian
\begin{comment}
\begin{align}\label{eq:Hamlocal}
    H=\sum_n \sigma^x_n\sigma^x_{n+1}+\sigma^y_n\sigma^y_{n+1}+\sigma^z_n\sigma^z_{n+1}+\sigma_n^x+h_n\sigma^z_n,
\end{align}
\end{comment}
\begin{align}\label{eq:Hamlocal}
    H=\sum_n J_n\sigma^z_n\sigma^z_{n+1}+\sigma_n^x+h_n \sigma^z_n,
\end{align}
with $h_n$ drawn uniformly from the interval $[-2,2]$ and $J_n$ drawn uniformly from the interval $[0.8,1.2]$. Here $\sigma_n^x$ and $\sigma_n^z$ are Pauli matrices acting at site $n$ and the sum runs over $L$ sites with either open or periodic boundary conditions.  This model has been studied previously at strong disorder, in the context of a rigorous proof of the existence of many-body localization~\cite{Imbrie_2016}. At the disorder strength considered here, however, it is known to thermalise~\cite{Abanin_2021}.

The SFF for this model is shown in Fig.~\ref{fig:Localmodel}. As expected from our results at large $N$ and small $\lambda$, the SFF is larger at intermediate times in systems with open boundary conditions than in systems with periodic boundary conditions, and this effect becomes more pronounced with increasing system size. In other circumstances one might dismiss a difference in behaviour between systems with different boundary conditions as an uninteresting finite-size effect. We do not believe the differences should be seen in that way here, for two reasons. First, the differences in behaviour according to the boundary conditions are of exactly the kind expected on theoretical grounds, and second, these differences grow larger with increasing system size, whereas other finite-size effects typically become smaller.  On this basis, we think that these data are evidence that the results we have obtained at large $N$ are relevant more generally.

We have also examined the consequences of reducing the level of disorder by drawing $h_n$ from the narrower interval $[-1,1]$. In this case, an enhancement of the SFF for open boundary conditions persists, but is not as large as in Fig.~\ref{fig:Localmodel}. Since reduced disorder implies a larger value of the energy diffusion constant, this change is consistent at a qualitative level with our results at large $N$. More specifically, we find in Eq.~\eqref{eq:wlate} that in increase in $D$ leads to a decrease in $w(t)$ at fixed $t$, which in turn implies an earlier onset of the ramp in the SFF. 

While we are not aware of a previous study of the influence of a change in boundary conditions on the SFF for a model with a time-independent Hamiltonian, data in Ref.~\cite{Kos_2020} shows an (unremarked) intermediate-time peak in the SFF: see especially the connected SFF, shown in Fig.~5(a) of that paper. In addition, results from simulations with moderately weak intersite coupling, described in Ref.~\cite{altland2025pathintegralapproachquantum} (which appeared while the current paper was in preparation) show similar effects for small $N$. For Floquet models, an equivalent dependence on boundary conditions and on system size to the one we discuss here is displayed in Fig.~5 of Ref.~\cite{Garratt_2021}. 

\section{Summary and outlook}\label{sec:summary}

In this work we have studied energy transport and spectral correlations in a minimal model of a spatially extended, chaotic many-body quantum system with local interactions. Our treatment, which is exact in the limits of large local Hilbert space dimension and weak intersite coupling, has three main steps. One is to show that energy dynamics in the model is described by a classical master equation. The second is to recognise that the solution to this master equation at a given time can be decomposed into separate contributions, according to whether energy has been exchanged between each neighbouring pair of sites during the evolution up to that time. The third step is to relate this decomposition to spectral correlations.

The outcome of this analysis is that (at times later than the onset of the ramp in the SFF for a single site) the SFF for a system of $L$ sites is given by a sum of terms varying as $t^P$, with integer $P$ in the range $1\leq P \leq L$. Each term has a weight that depend on $P$ and $t$. At early times, large values of $P$ dominate, while at sufficiently late times only $P=1$ is important. As a consequence, the SFF at early times in a large system is much larger than would be expected from random matrix theory. It has a crossover to random matrix behaviour at long times, controlled by three timescales. One is the timescale $\lambda^2t \sim 1$ for the onset of energy transfer between neighbouring sites; a second is the timescale $\lambda^2 t \sim (\ln L)^2$ for energy to have been exchanged between all neighbouring pairs of sites; and the third is the time for equilibration of the many-body energy distribution, which varies as $L^2/D$.

We have tested our conclusions using numerical simulations in two ways. For a two-site system, exact diagonalisation is possible with $N$ large. In this case we find excellent quantitative agreement with our analytical results. For larger $L$, exact diagonalisation is possible only for small $N$ and so we do not expect to make a similar quantitative comparison. Nevertheless, we find the same qualitative features at $N=2$ that we expect from our large $N$ calculations. We therefore believe that our approach gives a perspective on spectral correlations that is of general relevance. 

Our calculations offer a description of the dynamics of the many-body system in Fock space that has two important features. One of these is that it is conserving, in the sense that the probability density associated with the many-body wavefunction is conserved. While this is an elementary requirement, it is also a strong constraint on approximation schemes. For this reason, we think that our approach provides a suitable approximation outside the large $N$, small $\lambda$ limit, and that alternatives may be difficult to construct. The second feature is that contributions to dynamics are expressed in terms of paired Feynman paths. The weight for such a pair is the square modulus of the quantum amplitude; it is therefore positive and can be interpreted as a classical probability, providing the basis for a master equation.

There are some close parallels between our treatment of many-body dynamics in Fock space, and the established theory of diffusion in single-particle models of disordered conductors \cite{Edwards_1958}. For the latter, the standard approach is known as the diffusion approximation. It is conserving and can be viewed in terms of paired Feynman paths, in this case in real space \cite{Khmelnitskii_1985}. There are multiple routes to the diffusion approximation in single-particle models. The one that is closest to ours is via Wegner's $n$-orbital model \cite{Wegner_1976}, in which there is an $n$-dimensional Hilbert space associated with each site of a lattice, but with the Hilbert space for the entire system generated as a direct sum of single-site contributions, rather than the direct product we use here. %Eq.~\eqref{eq:Hamiltonian}. 
In this case, too, paired Feynman paths are exact for $n$ large. Alternatively, essentially the same calculations can be justified when disorder is weak, using the inverse of the dimensionless conductance as an expansion parameter \cite{Altshuler_85,Akkermans_2007}. It would be interesting to explore other possible justifications of the many-body calculations we have presented, besides large $N$ and small $\lambda$: a possible approach might be to combine several sites, each with small local Hilbert space dimension, into effective sites with large local Hilbert space dimension that can be treated using our methods.

We hope that our detailed results derived from a microscopic model will provide a useful point of comparison in efforts \cite{Winer_2022,Galitski_2022,altland2025pathintegralapproachquantum} to construct field-theoretic descriptions of many-body quantum dynamics, and several parallels are immediately evident. First, the role of diffusion in controlling the late-time approach of the SFF to a ramp emerges from energy conservation in a hydrodynamic approach to the SFF \cite{Winer_2022} and in a very recent path integral approach \cite{altland2025pathintegralapproachquantum}, as it does in our calculations and in an earlier one for a Floquet quantum circuit with a conserved density \cite{Friedman_2019}. Second, the importance of intersite coupling for the form of the SFF at intermediate times is captured within the path integral approach \cite{altland2025pathintegralapproachquantum} by a mechanism broadly similar to the one that operates in our calculations, again mirroring behaviour in Floquet quantum circuits \cite{Chan_PRL}. Third, the idea that the ergodic phase is characterised by spontaneous symmetry breaking \cite{GarrattPRL_2021} is central to the sigma-model calculations of Ref.~\cite{Galitski_2022}, and is represented in our approach by coupling of all neighbouring pairs of sites via energy exchange at late times.

There are naturally some aspects of our results that involve microscopic details and so are unlikely to be reproduced by an effective theory. One is the constant of proportionality in the relation $K(t) \propto t$
that holds on the ramp of the SFF, given correctly in terms of the bandwidth of the many-body spectrum by our approach. Another is the late-time dependence of the coupling between sites, represented in our calculations by the function $w(t)$.  

For future work, an important direction will be to extend our techniques to a treatment of operator spreading and entanglement dynamics. In general, calculations of dynamics in many-body systems require knowledge of matrix elements of $(e^{i{\cal H}t}\otimes e^{-i{\cal H}t})^{\otimes m}$ for positive integer $m$. Here we have considered only the case $m=1$, while the extensions would require $m\geq 2$.

\section*{Acknowledgements}
One of us (JTC) thanks Amos Chan, Andrea De Luca and Sam Garratt for previous collaborations on related work. Both authors thank Benedickt Placke and Curt von Keyserlingk for helpful discussions. We acknowledge support from the UK EPSRC under grant EP/X030881/1, and from a Leverhulme Trust International Professorship grant (Award Number: LIP-2020-014), funding a Leverhulme-Peierls Fellowship for DH.

\appendix 

\section{Correlations on the scale of the Heisenberg time}\label{sec:crossover}

We require information about correlations on the scale of the Heisenberg time at two points in our calculations: in the derivation of Eq.~\eqref{eq:fully}; and to generate the very late time form of the curves appearing in Fig.~\ref{fig:NumericsSFF}~(b) and (d). This information is contained in standard results from random matrix theory, but we are not aware of a previous publication that presents it in the form that we need. We therefore give a derivation in this Appendix. We first set out the ideas for a single-site contribution $H_n$ to $\cal H$, then indicate how they extend to a bigger system.

For the derivation of Eq.~\eqref{eq:fully} we need to simplify expressions of the form
\begin{equation}\label{eq:appendix1}
    f(t) =   \left[ \sum_{\varepsilon_1,\,\varepsilon_2} F(\epsilon_1,\epsilon_2)\, e^{it(\varepsilon_1 - \varepsilon_2)}\right]_{\rm av}
\end{equation}
at times $t$ of order the Heisenberg time. Here, $\varepsilon_1$ and $\varepsilon_2$ are two eigenvalues of $H_n$ (and so in this Appendix, but not elsewhere, we are depart from the convention we introduced in Eq.~\eqref{eq:basis}, that the subscript $n$ on $\varepsilon_n$ refers to a site). In addition, $F(\varepsilon_1,\varepsilon_2)$ varies smoothly with $\varepsilon_1$ and $\varepsilon_2$, and the form of Eq.~\eqref{eq:appendix1} follows that of Eq.~\eqref{eq:singlesite}. We use as input the two-point correlation function of the energy level density, written in the form \cite{Mehta_Random_2004}
\begin{multline}\label{eq:DoScorr}
   N^{-2} [ ({\rm Tr} \,\delta(\varepsilon_1 - H_n)) ({\rm Tr} \, \delta(\varepsilon_2 - H_n))]_{\rm av}\\ = \rho(\varepsilon_1)\,\rho(\varepsilon_2) + \rho(\varepsilon)^2 g(s)\,,
\end{multline}
where the mean energy is $\varepsilon = (\varepsilon_1 + \varepsilon_2)/2$, the scaled energy separation is $s= N\rho(\varepsilon)(\varepsilon_1 - \varepsilon_2)$, and we assume $N$ large for fixed $s$. The connected part of the correlator is \cite{Mehta_Random_2004}
\begin{equation}
    g(s) = \delta(s) - \frac{\sin^2 \pi s}{(\pi s)^2}\,.
\end{equation}

For large $t$ only this connected part contributes to Eq.~\eqref{eq:appendix1} and we have
\begin{equation}
    f(t) = \int {\rm d}\varepsilon\, F(\varepsilon,\varepsilon)\, k(t,\varepsilon)
\end{equation}
with
\begin{eqnarray}\label{eq:Def-k}
    k(t,\varepsilon) &=& N \rho(\varepsilon) \int {\rm d}s \, g(s) \, e^{ist/(N\rho(\varepsilon))} \nonumber\\
    &=& \left\{ \begin{array}{ccc} 
    {|t|}/{2\pi}&\quad& |t| \leq 2\pi N \rho(\varepsilon)\phantom{\,.}\\ && \\N\rho(\varepsilon) & & |t| > 2\pi N \rho(\varepsilon)\,.
    \end{array}\right.
\end{eqnarray}

In our calculations for a two-site system, we use a similar approach to derive an analytical form for the interpolation between the ramp and the plateau in the SFF at the Heisenberg time $t\sim N^2$ for the two-site system. Our starting point is Eq.~\eqref{eq:SFFdecomp} with the form for $R_X(t)$ at long times given in Eq.~\eqref{eq:Rlongt}. This is exact in the time window $\lambda^{-2} \ll t \ll N^2$. We extend it to all times $\lambda^2 t \gg 1$ by writing the SFF as
\begin{equation}
    K(t) = \int {\rm d}\varepsilon^\prime_{\rm tot}\, k_{\rm tot}(t,\varepsilon^\prime_{\rm tot})
\end{equation}
with
\begin{equation*}
    k_{\rm tot}(t,\varepsilon_{\rm tot}) = \left\{\begin{array}{llr}
    t/2\pi &\,\, & 1 \ll t \leq t(\varepsilon_{\rm tot})
\\
    &&\\
     N^2 \rho_{\rm tot}(\varepsilon_{\rm tot}) && t> t(\varepsilon_{\rm tot})
    \end{array}
    \right.
\end{equation*}
where $t(\varepsilon_{\rm tot}) = 2\pi N^2 \rho_{\rm tot}(\varepsilon_{\rm tot})$. This is used to generate the analytical prediction shown in Fig.~\ref{fig:NumericsSFF} for $K(t)$ at times $\lambda^2 t \gg 1$.

\section{Analytical estimates for the diffusion constant}\label{sec:Diffusionimprovedestimate}
    \begin{figure}[t!]
    \centering
    \includegraphics[width=0.45 \textwidth]{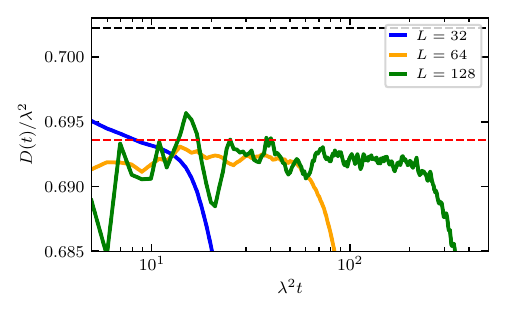}
    \caption{Time dependent diffusion constant  Second moment of the autocorrelator, $D(t)=\frac{1}{2t}\sum_n n^2 \braket{\varepsilon_n(t)\varepsilon_0(0)}$.  Comparison with the analytical estimate for the diffusion equation $D=0.7022 \lambda^2$ Eq.~\eqref{eq:Dapprox}~(black dashed line) and the improved estimate in $D=0.6936 \lambda^2$~Eq.~\eqref{eq:Dnew}. Extending the ansatz as in Eq.~\eqref{eq:Ansatzone} significantly improves agreement with numerical results.}
    \label{fig:Diffusionfinitesize}
    \end{figure}

    In this appendix, we show how a systematic extension of the ansatz described below Eq.~\eqref{eq:Ansatzzerothorder} reduces the tension between the numerical and analytical results for the diffusion constant $D$.

    To do so, we add higher-order moments of the energy to Eq.~\eqref{eq:Ansatzzerothorder}. For symmetry reasons, even moments in the energy do not change the estimate for the diffusion constant. The next possible extension is therefore obtained by adding local cubic terms of the form $\varepsilon_n^3$, $\varepsilon_n^2\varepsilon_{n+1}$, and $\varepsilon_n^2\varepsilon_{n-1}$. 
    This leads to an ansatz for the energy distribution at a given time, of the form:
    \begin{multline}\label{eq:Ansatzone}
		\frac{1}{Z_1}\prod_n \rho(\varepsilon_n) e^{-\beta_n \varepsilon_n- \delta_{n^3} \varepsilon_n^3}\\\times e^{-\delta_{n^2(n+1)} \varepsilon_n^2\varepsilon_{n+1}-\delta_{n^2(n-1)} \varepsilon_n^2\varepsilon_{n-1}}.
	\end{multline}
    The terms $\beta_n$, $\delta_{n^3}$, $\delta_{n^2(n+1)}$ and $\delta_{n^2(n-1)}$ are Lagrange multipliers and $Z_1$ normalizes the probability distribution. 

    Assuming that the amplitude modulations and the Lagrange multipliers are small, Eq.~\eqref{eq:Ansatzone} can be expanded to linear order in the Lagrange multipliers to obtain expectation values $\braket{\varepsilon_n}$, $\braket{\varepsilon_n^3}$, $\braket{\varepsilon_n^2\varepsilon_{n+1}}$, and $\braket{\varepsilon_n^2\varepsilon_{n-1}}$.
    This gives the following expressions:
    \begin{subequations}\label{eq:Lagrangemultiliers}
		\begin{align}
            \nonumber
			\braket{\varepsilon_n}&=\beta_n+2\delta_{n^3}\\&+\delta_{(n-1)^2 n}+\delta_{(n+1)^2 n}\\\nonumber
            \braket{\varepsilon_n^3}&=2\beta_n+4\delta_{n^3}+2\delta_{(n-1)^2 n}\\&+2\delta_{(n+1)^2 n
            }\\\nonumber
            \braket{\varepsilon_n^2 \varepsilon_{n+1}}&=\beta_{n+1}+2\delta_{(n+1)^3}+2\delta_{n^2 (n+1)}\\&+\delta_{(n+2)^2(n+1)}\\\nonumber
            \braket{\varepsilon_{n}^2\varepsilon_{n-1}}&=\beta_{n-1}+2\delta_{(n-1)^3}\\&+\delta_{n^2 (n-1)}+2\delta_{(n-2)^2(n-1)}
		\end{align}
	\end{subequations}

    Inserting the ansatz Eq.~\eqref{eq:Ansatzone}
    into the master equation Eq.~\eqref{eq:master} and expanding to linear order in the Lagrange multipliers leads to a translation-invariant evolution equation for the moments $\braket{\varepsilon_n}$,$\braket{\varepsilon_n^3}$, $\braket{\varepsilon_n^2\varepsilon_{n+1}}$, and $\braket{\varepsilon_n^2\varepsilon_{n-1}}$ as a function of the Lagrange multipliers. Transforming these equations to momentum space and inverting the expression Eq.~\eqref{eq:Lagrangemultiliers} gives a linear homogeneous differential equation for the energy moments.

    Defining the four-dimensional vector
    ${\vec{c}(n,t)=(\varepsilon_n,\varepsilon_n^3,\varepsilon_n^2\varepsilon_{n+1},\varepsilon_n^2\varepsilon_{n-1}})$
    and defining the coefficients of the matrix $\mathbf{C}(k,\omega)$ as
    \begin{align}\label{eq:Definition1}
    \begin{split}
		[\mathbf{C}(k,\omega)]_{ij}=\sum_n \int \rd t \,\theta(t) &\,\langle \vec{c}_i(n,t)\vec{c}_j(0,0)\rangle\times\\& \times\exp(i(kn-\omega t)),
    \end{split}
	\end{align}
    the differential equation can be expressed in terms of a matrix equation
	\begin{align}
		(-i\omega -\mathbf{G}_0(k))\mathbf{C}(k,\omega)=\mathbf{A}(k),
	\end{align}
where $\mathbf{G}_0(k)$ is obtained by the Fourier transform of the equations of motion. While the explicit expression is a bit tedious to obtain, we note that all coefficients of this matrix can be expressed as explicit integrals and evaluated numerically.

    $\mathbf{A}(k)$ is obtained from the correlations at time $t=0$. At lowest order in the Lagrange multipliers, these correlations are identical with the infinite temperature average and given by
    \begin{align}
    \mathbf{A}(k)=\begin{pmatrix}
    1&2&e^{ik}&e^{-ik}\\
    2&5&2e^{ik}&2e^{-ik}\\
    e^{-ik}&2e^{-ik}&2&4\\
    e^{ik}&2e^{ik}&4&2
    \end{pmatrix}
    \end{align}

    Within this ansatz, the estimate for the diffusion constant is obtained as  
	\begin{align}\label{eq:Dnew}D=\lim_{k\rightarrow0}\lim_{\omega\rightarrow0}\left(\frac{1}{k^2 \mathbf{C}(k,\omega)}\right)_{1,1}\approx 0.6936 \lambda^2.
	\end{align}
    The last result is obtained by numerical evaluation of the $(4\times 4)$ matrix $C(k,\omega )$ for $k,\omega \rightarrow \infty.$
    To compare the improved estimate with numerics, we obtain an estimate for the diffusion constant by evaluating the variance of the autocorrelator as
    \begin{equation}
        D(t)=\frac{1}{2t}\sum_n n^2 \braket{\varepsilon_n(t)\varepsilon_0(0)}.
    \end{equation}
    The results are shown in Fig.~\ref{fig:Diffusionfinitesize}.
    The refined estimate in Eq.~\eqref{eq:Dnew} yields substantially improved agreement with the numerical data. Moreover, the ansatz in Eq.~\eqref{eq:Ansatzone} admits systematic extensions, for example by incorporating higher-order local moments. This gives a rather complete description of the late-time energy transport governed by Eq.~\eqref{eq:master}.
    
 \subsection{Estimates for the diffusion constant at finite temperature}
 
     \begin{figure}[t!]
    \centering
    \includegraphics[width=0.45 \textwidth]{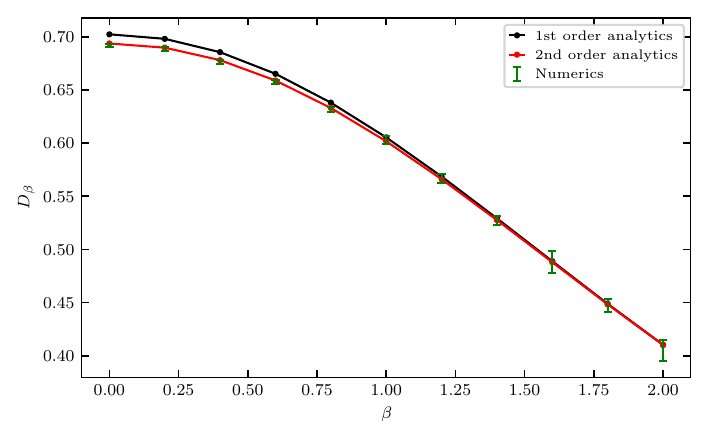}
    \caption{Numerical results for the diffusion constant $D_\beta$ as a function of inverse temperature $\beta$~(green error bars). Comparison with the estimate of Eq.~\eqref{eq:Dbeta}~(black line) and an improved estimate analogous to the one described in 
    Appendix~\ref{sec:Diffusionimprovedestimate} (red line). With increasing inverse temperature $\beta$, the agreement with Eq.~\eqref{eq:Dbeta} improves.}
    \label{fig:Diffusionbeta}
    \end{figure}
Using the same steps as above, the estimate for the diffusion constant can be improved for finite temperatures. In Fig.~\ref{fig:Diffusionbeta}, we present data for the diffusion constant as a function of the inverse temperature $\beta$ and a comparison with approximate analytical results. The agreement is excellent. Furthermore, the simplest approximation, Eq.~\eqref{eq:Dbeta}, becomes more accurate with increasing $\beta$, suggesting that it is enough to evaluate Eq.~\eqref{eq:Dbeta} in order to obtain the low temperature limit $\beta\rightarrow \infty$.

To evaluate Eq.~\eqref{eq:Dbeta} in the low-temperature limit $\beta\to\infty$, we expand the energy integrals around the lower band edge, $\varepsilon_m=-2$. We write  $u_m = \varepsilon_m +2$ and $u = \varepsilon_{\rm tot} +4$. This gives at leading order in $u_m$ and $u$ $\rho(\varepsilon_m) = \sqrt{u_m}/\pi$ and $\gamma(\varepsilon^\prime_{\rm tot}) = u^2_{\rm}\lambda^2$. We obtain 
\begin{align}
Z_\beta\simeq e^{2\beta} \int_0^\infty \rd u_m \sqrt{u_m}\pi e^{-\beta u_m}=\frac{1}{2\pi^{1/2}\beta^{3/2}}e^{2\beta}
\end{align}
for the partition function. Similarly, we find
\begin{align}
\chi_\beta^2\simeq\frac{3}{2\beta^2}
\end{align}
and
\begin{align}
\begin{split}
&\int {\rm d}\varepsilon_1\,{\rm d}\varepsilon_2\,{\rm d}\varepsilon_3\,
\rho(\varepsilon_1)\rho(\varepsilon_2)\rho(\varepsilon_3)\rho(\varepsilon_1+\varepsilon_2-\varepsilon_3)
 \\&\times 
e^{-\beta(\varepsilon_1+\varepsilon_2)}
\left(\varepsilon_1^2-\varepsilon_1\varepsilon_3\right)\simeq \frac{45}{64 \pi^2}\frac{e^{4\beta}}{\beta^7}. 
\end{split}
\end{align}
Combining all factors recovers the result given  in Eq.~\eqref{eq:Dbeta1}.
       
\bibliography{references}

\end{document}

%% file: preamble.tex
\usepackage[utf8]{inputenc}
\usepackage{fullpage}
\usepackage{graphicx}
\usepackage{amsmath}
\usepackage{amsthm}
\usepackage{enumitem}
\usepackage{amssymb}
\usepackage{wasysym}
\usepackage{oplotsymbl}
\usepackage{bm} % bold italic vectors
\usepackage{dsfont} % identity operator
\usepackage{mathrsfs} % mathscr font
\usepackage{physics}
\usepackage{mathtools}
\usepackage{quantikz}
\usepackage{braket}

\usepackage[dvipsnames]
{xcolor}
\definecolor{darkblue}{rgb}{0,0,.65}
\definecolor{darkgreen}{rgb}{0.28,0.41,0.19}
\definecolor{nicegreen}{rgb}{0.28,0.85,0.19}

\usepackage[colorlinks=true,linkcolor=blue,allcolors=blue]{hyperref}
\usepackage[capitalize,nameinlink]{cleveref}

\usepackage{comment}

\def\equationautorefname~#1\null{Eq. (#1)\null}

\newcommand{\appref}[1]{\hyperref[#1]{App.~\ref*{#1}}}

\renewcommand\vec{\bm}
\newcommand{\rd}{\mathrm{d}}